\documentclass{elsart}

\usepackage{amsbsy}
\usepackage{amssymb}
\usepackage{bm}
\usepackage{epsfig}
\usepackage{subfigure}
\usepackage{verbatim}
\usepackage{dsfont}

\begin{document}
\begin{frontmatter}
\title{Fluctuating force-coupling method for simulations of colloidal suspensions}
\author{Eric E. Keaveny}
\ead{e.keaveny@imperial.ac.uk}
\address{Department of Mathematics, Imperial College London, South Kensington Campus, London, SW7 2AZ, UK}
\begin{abstract}
The resolution of Brownian motion in simulations of micro-particle suspensions can be crucial to reproducing the correct dynamics of individual particles, as well as providing an accurate characterisation of suspension properties.  Including these effects in simulations, however, can be computationally intensive due to the configuration dependent random displacements that would need to be determined at every time step.  In this paper, we introduce the fluctuating force-coupling method (FCM) to overcome this difficulty, providing a fast approach to simulate colloidal suspensions at large-scale.  We show explicitly that by forcing the surrounding fluid with a fluctuating stress and employing the FCM framework to obtain the motion of the particles, one obtains the random particle velocities and angular velocities that satisfy the fluctuation-dissipation theorem.  This result holds even when higher-order multipoles, such as stresslets, are included in the FCM approximation.  Through several numerical experiments, we confirm our analytical results and demonstrate the effectiveness of fluctuating FCM, showing also how Brownian drift can be resolved by employing the appropriate time integration scheme and conjugate gradient method. 
\end{abstract}
\maketitle
\end{frontmatter}
\section{Introduction}

Brownian motion, or the random movement of particles suspended in liquid \cite{Russel1981}, results from the many collisions between the particles and the molecules that make up the surrounding fluid.  While this is inherently linked to the discrete, molecular nature of the fluid, the effects of Brownian motion extend upwards to longer, continuum length scales, affecting not only the dynamics of individual particles, but also the properties of suspensions themselves.   For example, Brownian motion is known to affect the rheological properties of particulate suspensions, changing their linear response to applied stresses, as well as contributing to their non-Newtonian behaviour \cite{Batchelor1977,Foss2000,Cheng2002}.  In biological systems, the diffusion of Brownian particles is a fundamental mechanism of transport, regulating rates of many life processes, especially those in crowded intracellular environments \cite{Grima2010}.  Characterising and quantifying the role of Brownian motion in these contexts where interparticle forces, hydrodynamic interactions, and geometric constraints play a strong role presents a current computational challenge.  Moreover, with the development of particle self-assembly and aggregation-based fabrication techniques \cite{Whitesides2002,Glotzer2004}, as well as the increasing number of highly engineered active, flow-generating and field-responsive micro-particles \cite{Paxton2004,Palacci2013,Walther2008}, accurately characterising the effects of Brownian motion on suspension dynamics and structure is of fundamental technological importance.  

In simulation techniques such as Brownian dynamics \cite{Ermak1978} and Stokesian dynamics \cite{Brady1988}, Brownian motion is incorporated by introducing random particle velocities at each time step.  However, in order for either of these methods to yield the correct particle diffusion, the random particle velocities must follow precise statistics, where their correlations are proportional to the hydrodynamic mobility matrix \cite{Einstein,Batchelor1976}.  This requires one to compute the square root of the mobility matrix, an $\mathcal{O}(N^3)$ calculation, at every time step.   Thus, including the effects of Brownian motion adds significant computational overhead to both of these methods, and as a result, has limited such simulations to two extreme cases -- small particle numbers with the hydrodynamic interactions adequately resolved, or larger-scale simulations in which the hydrodynamic interactions are ignored completely.  Further, the multiplicative noise, or noise whose amplitude depends on the particle positions, introduced by the hydrodynamic interactions yields also a Brownian drift term \cite{Ermak1978} that is proportional to the divergence of the mobility matrix.  This term also needs to be computed in order to produce the correct particle dynamics.

One approach to overcoming these limitations is to utilise a polynomial expansion of the matrix square root \cite{Fixman1986}.   This method has been used successfully in conjunction with both Brownian and Stokesian dynamics \cite{Banchio2003,Jendrejack2000}, allowing for simulations with significantly more particles than would otherwise be possible.  Another approach to increase the speed of Brownian simulations, and the one that we will pursue in this work, is to introduce a white-noise, fluctuating stress \cite{LandauLifshitz} to drive the surrounding fluid and require that the resulting velocity field satisfy the no-slip condition, or some approximation to it, on the particle surfaces.  Indeed, Fox and Uhlenbeck \cite{Fox1970} showed for rigid particles that this approach does yield the correct particle velocity correlations and, consequently, the correct diffusion matrix for the suspension.  Since the fluctuating stress itself is \emph{independent} of the particle configuration, the $\mathcal{O}(N^3)$ matrix square root computation is not required.  While this approach does require one to solve for the random fluid flow, such fluid flow computations are typically already performed to find the deterministic motion of the particles.  The effectiveness of fluctuating stresses in resolving Brownian motion has been demonstrated in a variety of simulation techniques.  They have been successfully employed in large-scale Lattice-Boltzmann simulations of particulate suspensions \cite{Ladd1993,Ladd1994,Ladd1994II,Ladd2001}, as well as more traditional, continuum mechanics based simulations of Brownian particles and structures.  Fluctuating stresses have been used with the distributed Lagrange multiplier (DLM) method \cite{Sharma2004} where the induced fluctuating flow is constrained at the grid points within the solid particle.  Recently, they have been successfully employed with immersed-boundary \cite{Peskin2002,Atzberger2007,Atzberger2011} and ``blob'' methods \cite{Balboa2013}, resolving the fluctuations of flexible structures, even in cases where inertial effects are present and lead to power-law tails in the time-correlations of the particle velocities \cite{Hinch1975,Russel1981}.  

Based on the success of these approaches, we utilise fluctuating stresses, the fluid flows they produce, and the simulation technique known as the force-coupling method (FCM) to develop a fast method for large-scale simulations of suspensions of interacting particles.  FCM \cite{Maxey2001,Lomholt2003,Dance2003,Yeo2010} employs regularised multipole expansions of the force distributions the particles exerts on the surrounding fluid and spatial averaging of the resulting flow to obtain the particle motion.  It has been shown to be very effective for large-scale simulations of suspensions and particle-laden flows \cite{Xu2002,Pivkin2006,Climent2004,Yeo2010jfm} over a wide range of volume fractions.  Here, we show analytically that when the surrounding fluid is also forced by a fluctuating stress, FCM yields random particle velocity and angular velocity correlations consistent with the fluctuation-dissipation theorem \cite{Russel1981}.  A main result of this work is that fluctuating FCM gives the proper correlations even when higher-order multipoles, such as the rotlet and stresslet, are included in the multipole expansion.  We provide numerical examples confirming these results.  In addition, for dynamic fluctuating FCM simulations, we show how to recover Brownian drift using Fixman's midpoint time integration scheme \cite{Fixman1978,Grassia1995} and the conjugate gradient method.  We employ this scheme to examine long-time diffusion of interacting particles and suspension dynamics in cellular flow fields.

\section{Particle motion}
In this study, we will be considering a suspension of $N$ rigid spherical particles, each having radius $a$.  Each particle $n$, $(n = 1,\dots, N)$, is centred at $\mathbf{Y}_n$ and can be subject to external forces $\mathbf{F}_n$, and external torques $\bm{\tau}_n$.  We will be considering the motion of these particles in the over-damped, or Brownian dynamics \cite{Ermak1978}, limit where the Reynolds number \cite{KimBook,BatchelorBook} is low, and fluid and particle inertia are neglected.  While working in this limit does not resolve the power-law decay of the velocity autocorrelation function, it provides an accurate description of diffusive motion for times $t \gg \rho a^2/\eta$ ($\rho$ is the density of the fluid and $\eta$ the shear viscosity) \cite{Russel1981}, making it appropriate for describing the dynamics of suspensions of micron-scale, colloidal particles.  In this limit, the equations of motion can be written as
\begin{equation}
\frac{d\mathcal{Y}}{dt} = \mathcal{V} + \tilde{\mathcal{V}} + k_BT\mathbf{\nabla}_{\mathcal{Y}} \cdot \mathcal{M}^{\mathcal{VF}}
\label{eq:PM1} 
\end{equation}
where $\mathcal{Y}$ is the $3N \times 1$ vector containing the components of $\mathbf{Y}_n$ for all of the particles, $\mathcal{V}$ holds the components of the deterministic particle velocities, and $\tilde{\mathcal{V}}$ gives the random velocities of the particles due to Brownian motion.  The Brownian drift term is given by $k_BT\mathbf{\nabla}_{\mathcal{Y}} \cdot \mathcal{M}^{\mathcal{VF}}$ where $k_B$ is Boltzmann's constant, $T$ is the temperature of the system, and $\mathcal{M}^{\mathcal{VF}}$ is the translational mobility matrix as described below.  

The deterministic velocities, $\mathcal{V}$, as well as the particle angular velocities, $\mathcal{W}$, are given by
\begin{equation}
\left[\begin{array}{c}
\mathcal{V} \\
\mathcal{W} \\
\end{array}\right]=
\left[\begin{array}{cc}
\mathcal{M}^\mathcal{VF} & \mathcal{M}^\mathcal{VT} \\
\mathcal{M}^\mathcal{WF} & \mathcal{M}^\mathcal{WT}  \\
\end{array}\right]
=\mathcal{M}
\left[\begin{array}{c}
\mathcal{F} \\
\mathcal{T} \\
\end{array}\right]
\label{eq:PM1b}
\end{equation}
where $\mathcal{F}$ is the $3N\times 1$ vector containing the components of $\mathbf{F}_n$ for all $N$ particles, and $\mathcal{T}$ holds the components of $\bm{\tau}_n$.  The $6N\times 6N$ matrix $\mathcal{M}$ is the complete low Reynolds number mobility matrix and is comprised of the four $3N\times 3N$ submatrices $\mathcal{M}^\mathcal{VF}$, $\mathcal{M}^\mathcal{VT}$, $\mathcal{M}^\mathcal{WF}$, and $\mathcal{M}^\mathcal{WT}$.  The exact values of the mobility matrix entries are found by considering the Stokes equations
\begin{eqnarray}
	-\bm{\nabla} p + \eta \nabla^2\mathbf{u}&=&\mathbf{0} \nonumber\\
	\bm{\nabla} \cdot \mathbf{u} &=& 0
	\label{eq:PM2}
\end{eqnarray}
for fluid velocity $\mathbf{u}$ and pressure $p$ subject to the no-slip boundary conditions, $\mathbf{u} = \mathbf{V}_n + \bm{\Omega}_n\times(\mathbf{x} - \mathbf{Y}_n)$, on the surface of each particle, where the velocity, $\mathbf{V}_n$, and angular velocity, $\bm{\Omega}_n$, for each particle $n$ are unknown.  By solving the Stokes equations with the additional conditions that $\bm{\tau}_n = \mathbf{0}$ for all $n$, $\mathbf{F}_m = \mathbf{e}_i$ ($i = 1, 2,$ or $3$), and $\mathbf{F}_n = \mathbf{0}$ for $n\neq m$, the resulting values of $\mathbf{V}_n$ will give the $3(m-1) + i$ column of $\mathcal{M}^\mathcal{VF}$ while the values of $\bm{\Omega}_n$ are the $3(m-1) + i$ column of $\mathcal{M}^\mathcal{WF}$.  If instead, we take $\mathbf{F}_n = \mathbf{0}$ for all $n$, but $\bm{\tau}_m = \mathbf{e}_i$ and $\bm{\tau}_n = \mathbf{0}$ for $n\neq m$, the values of $\mathbf{V}_n$ will be the $3(m-1) + i$ column of $\mathcal{M}^\mathcal{VT}$, while the values of $\bm{\Omega}_n$ are the $3(m-1) + i$ column of $\mathcal{M}^\mathcal{WT}$.

The remaining two terms on the right hand side of Eq. (\ref{eq:PM1}) are due to Brownian motion.  The focus of this paper is how to provide a consistent approximation of these terms using fluctuating FCM.  In order to achieve the correct particle diffusion for a suspension, the statistics of random particle velocities, $\tilde{\mathcal{V}}$, as well as the random particle angular velocities, $\tilde{\mathcal{W}}$, must satisfy a precise relation known as the fluctuation-dissipation theorem \cite{Russel1981}.  The fluctuation-dissipation theorem states that 
\begin{eqnarray}
\langle \tilde{\mathcal{V}}(t)\rangle &=& 0\\
\langle \tilde{\mathcal{W}}(t)\rangle &=& 0\\
\left\langle \left[\begin{array}{c}
\tilde{\mathcal{V}}(t) \\
\tilde{\mathcal{W}}(t) \\
\end{array}\right] 
\left[\begin{array}{cc}
\tilde{\mathcal{V}}^T(t') & \tilde{\mathcal{W}}^T(t')
\end{array}\right]
\right\rangle &=& 2k_BT \mathcal{M}\delta(t-t')
\label{eq:PM2}
\end{eqnarray}
where we have used $\langle \cdot \rangle$ to denote the ensemble average of a quantity.  While the random velocities and angular velocities have zero mean, the correlations depend directly on the mobility matrix, $\mathcal{M}$.  We will show that fluctuating FCM yields random velocities and angular velocities that satisfy this relationship, with the correlation matrix being the FCM approximation to the mobility matrix.

The second term introduced by Brownian motion is Brownian drift, $k_BT\mathbf{\nabla}_{\mathcal{Y}} \cdot \mathcal{M}^{\mathcal{VF}}$.  This drift corresponds to the mean particle velocities established during the inertia-friction relaxation time-scale ($t \ll m/(6 \pi \eta a$)) not resolved in the over-damped limit.  It can be derived by considering small displacements of the particles in the full Langevin equations and carefully taking the limit $6\pi \eta a t/m \rightarrow \infty$ \cite{Grassia1995}.  In this work, we show that for dynamic fluctuating FCM simulations, a direct computation of the Brownian drift term can be avoided by employing the midpoint time integration scheme developed by Fixman \cite{Fixman1978,Grassia1995}.  To use this scheme, however, one must work with random forces, $\tilde{\mathcal{F}}$ and torques, $\tilde{\mathcal{T}}$, rather than $\tilde{\mathcal{V}}$ and $\tilde{\mathcal{W}}$ that fluctuating FCM outputs.  We show, however, that these random forces and torques can be found using the conjugate gradient method, allowing for the dynamics of colloidal suspensions to be resolved in an efficient manner.

\section{The force-coupling method} \label{sec:FCM}

FCM provides an accurate and efficient way of simulating the deterministic motion of particles in dilute suspensions.  With respect to Eq. (\ref{eq:PM1b}), it corresponds to determining the velocities and angular velocities using an approximation of the mobility matrix.  We provide here an overview of FCM, summarising the results presented in \cite{Maxey2001,Lomholt2003,Dance2003,Yeo2010} and establishing the formulation we will use in our analysis of fluctuating FCM.

In FCM, each particle is represented by a low order finite-force multipole expansion in the Stokes equations
\begin{eqnarray}
	-\bm{\nabla} p + \eta \nabla^2\mathbf{u}&=&-\sum_n \mathbf{F}_n \Delta_n(\mathbf{x})  - \frac{1}{2}\bm{\tau}_n \times \bm{\nabla} \Theta_n(\mathbf{x}) - \mathbf{S}_{n} \cdot \bm{\nabla}\Theta_n(\mathbf{x})\nonumber\\
	\bm{\nabla} \cdot \mathbf{u} &=& 0.
	\label{eq:FCM1}
\end{eqnarray}
where $\mathbf{F}_n$, $\bm{\tau}_n$, and $\mathbf{S}_{n}$ are, respectively, the force, torque, and stresslet associated with particle $n$.  The stresslets provide a higher-order representation of the flow field generated by the particles, and, as discussed below, are determined through a condition on the local rate-of-strain.  In Eq. (\ref{eq:FCM1}), we also have the two Gaussian envelopes that are used to project the particle forces onto the fluid,
\begin{eqnarray}
\Delta_n(\mathbf{x})&=&(2\pi\sigma_{\Delta}^2)^{-3/2}\textrm{e}^{-|\mathbf{x} - \mathbf{Y}_n|^2/2\sigma_{\Delta}^2} \nonumber\\
\Theta_n(\mathbf{x})&=&(2\pi\sigma_{\Theta}^2)^{-3/2}\textrm{e}^{-|\mathbf{x} - \mathbf{Y}_n|^2/2\sigma_{\Theta}^2}.
\label{eq:FCM2}
\end{eqnarray} 
The length scales $\sigma_{\Delta}$ and $\sigma_{\Theta}$ are related to the radius of the particles through $\sigma_{\Delta} = a/\sqrt{\pi}$ and $\sigma_{\Theta} = a/\left(6\sqrt{\pi}\right)^{1/3}$.  After solving Eq. (\ref{eq:FCM1}), the velocity, $\mathbf{V}_n$, angular velocity, $\bm{\Omega}_n$, and local rate-of-strain, $\mathbf{E}_n$, of each particle $n$ are determined from 
\begin{eqnarray}
\mathbf{V}_n&=&\int\mathbf{u}\Delta_n(\mathbf{x})d^3\mathbf{x} \label{eq:FCM3a}\\
\bm{\Omega}_n&=&\frac{1}{2}\int\left[\bm{\nabla}\times\mathbf{u}\right] \Theta_n(\mathbf{x})d^3\mathbf{x} \label{eq:FCM3b}. \\
\mathbf{E}_n&=&\frac{1}{2}\int \left[\bm{\nabla}\mathbf{u} + (\bm{\nabla}\mathbf{u})^T\right]\Theta_n(\mathbf{x})d^3\mathbf{x}.
\label{eq:FCM3c}
\end{eqnarray}
where the integration is performed over the volume occupied by the fluid.  For rigid particles, the stresslets are found by enforcing the constraint $\mathbf{E}_n = \mathbf{0}$ for each $n$.  This is equivalent to stating that the local rates-of-strain can do no work on the fluid \cite{Lomholt2003}.  In demonstrating the resolution of particle Brownian motion, we will consider the case where the stresslets are included in FCM, as well as the case where they are omitted ($\mathbf{S}_n = \mathbf{0}$).

While this description of FCM closely follows its implementation, in demonstrating that fluctuating FCM yields linear and angular velocities consistent with the fluctuation-dissipation theorem, we will utilise the equivalent matrix representation \cite{Yeo2010} of FCM
\begin{equation}
\left[\begin{array}{c}
\mathcal{V} \\
\mathcal{W} \\
\mathbf{0}
\end{array}\right]=
\left[\begin{array}{ccc}
\mathcal{M}^{\mathcal{V}\mathcal{F}}_{FCM} & \mathcal{M}^{\mathcal{V}\mathcal{T}}_{FCM}  & \mathcal{M}^{\mathcal{V}\mathcal{S}}_{FCM} \\
\mathcal{M}^{\mathcal{W}\mathcal{F}}_{FCM} & \mathcal{M}^{\mathcal{W}\mathcal{T}}_{FCM} & \mathcal{M}^{\mathcal{W}\mathcal{S}}_{FCM} \\
\mathcal{M}^{\mathcal{E}\mathcal{F}}_{FCM} & \mathcal{M}^{\mathcal{E}\mathcal{T}}_{FCM} & \mathcal{M}^{\mathcal{E}\mathcal{S}}_{FCM}
\end{array}\right]
\left[\begin{array}{c}
\mathcal{F} \\
\mathcal{T} \\
\mathcal{S}
\end{array}\right].
\label{eq:FCM5}
\end{equation}
This relates the linear and angular velocities for all $N$ particles to the forces, torques, and stresslets on the particles.  The $11N\times 11N$ matrix on the right-hand side is the FCM grand mobility matrix \cite{Yeo2010}.  We can derive expressions for the entries of its submatrices using the FCM Gaussian envelopes, Eq. (\ref{eq:FCM2}), and the Stokeslet, 
\begin{equation}
\mathbf{G}(\mathbf{x}) = \frac{1}{8 \pi \eta |\mathbf{x}|}\left(\mathbf{I} + \frac{\mathbf{x}\mathbf{x}}{|\mathbf{x}|^2}\right),
\label{eq:FCM6}
\end{equation}
the Green's function for the Stokes equations \cite{Pozrikidis}.  For example, the flow generated by the force on particle $m$ can be expressed as
\begin{equation}
\mathbf{u}_m(\mathbf{x}) = \int \mathbf{G}(\mathbf{x}-\mathbf{y})\mathbf{F}_m\Delta_m(\mathbf{y})d^3\mathbf{y}.
\label{eq:fcmfu}
\end{equation}
Then, using Eq. (\ref{eq:FCM3a}), the velocity of particle $n$ due to $\mathbf{u}_m$ will be given by
\begin{equation}
\mathbf{V}_n = \int \int \mathbf{G}(\mathbf{x}-\mathbf{y})\mathbf{F}_m\Delta_m(\mathbf{y})\Delta_n(\mathbf{x})d^3\mathbf{y}d^3\mathbf{x}.
\end{equation}
We see, therefore, that the entries of $\mathcal{M}^{\mathcal{V}\mathcal{F}}_{FCM}$ that relate the velocity of particle $n$ to the force on particle $m$ are
\begin{eqnarray}
\mathcal{M}^{\mathcal{V}\mathcal{F}}_{FCM;n m} &=&\int \int \Delta_n(\mathbf{x}) \mathbf{G}(\mathbf{x} - \mathbf{y}) \Delta_m(\mathbf{y})d^3\mathbf{x}d^3\mathbf{y}.
\label{eq:FCM8}
\end{eqnarray}
Using the same approach, we can find similar expressions for the entries of the other matrices in Eq. (\ref{eq:FCM5}).  We provide these expressions in Appendix \ref{sec:FCMapp}.  

From Eq. (\ref{eq:FCM5}), we can determine the FCM approximation to the $6N \times 6N$ mobility matrix, $\mathcal{M}$ in Eq. (\ref{eq:PM1b}).  We consider separately the cases where the stresslets are ignored and where they are included.  If they are ignored, we have immediately $\mathcal{S} = \mathbf{0}$, and the FCM mobility matrix is simply
\begin{equation}
\mathcal{M}_{FCM} = 
\left[\begin{array}{cc}
\mathcal{M}^{\mathcal{V}\mathcal{F}}_{FCM} & \mathcal{M}^{\mathcal{V}\mathcal{T}}_{FCM}  \\
\mathcal{M}^{\mathcal{W}\mathcal{F}}_{FCM} & \mathcal{M}^{\mathcal{W}\mathcal{T}}_{FCM}
\end{array}\right]
\end{equation}
If the particle stresslets are included in FCM, they must be determined.  This can be done using the last line of Eq. (\ref{eq:FCM5}), which gives 
\begin{equation}
\mathcal{S} = -\mathcal{R}^{\mathcal{E}\mathcal{S}}_{FCM}\left(\mathcal{M}^{\mathcal{E}\mathcal{F}}_{FCM}\mathcal{F} + \mathcal{M}^{\mathcal{E}\mathcal{T}}_{FCM}\mathcal{T}\right).
\end{equation}
where we have written $\mathcal{R}^{\mathcal{E}\mathcal{S}}_{FCM} = (\mathcal{M}^{\mathcal{E}\mathcal{S}}_{FCM})^{-1}$.  From this expression for $\mathcal{S}$, we find the stresslet-corrected FCM mobility matrix is
\begin{equation}
\mathcal{M}_{FCM-S} = 
\left[\begin{array}{cc}
\mathcal{M}^{\mathcal{V}\mathcal{F}}_{FCM-S} & \mathcal{M}^{\mathcal{V}\mathcal{T}}_{FCM-S}  \\
\mathcal{M}^{\mathcal{W}\mathcal{F}}_{FCM-S} & \mathcal{M}^{\mathcal{W}\mathcal{T}}_{FCM-S}
\end{array}\right]
\end{equation}
where
\begin{eqnarray}
\mathcal{M}^{\mathcal{V}\mathcal{F}}_{FCM-S} &=& \mathcal{M}^{\mathcal{V}\mathcal{F}}_{FCM}-\mathcal{M}^{\mathcal{V}\mathcal{S}}_{FCM}\mathcal{R}^{\mathcal{E}\mathcal{S}}_{FCM}\mathcal{M}^{\mathcal{E}\mathcal{F}}_{FCM}, \\
\mathcal{M}^{\mathcal{V}\mathcal{T}}_{FCM-S} &=& \mathcal{M}^{\mathcal{V}\mathcal{T}}_{FCM}-\mathcal{M}^{\mathcal{V}\mathcal{S}}_{FCM}\mathcal{R}^{\mathcal{E}\mathcal{S}}_{FCM}\mathcal{M}^{\mathcal{E}\mathcal{T}}_{FCM}, \\
\mathcal{M}^{\mathcal{W}\mathcal{F}}_{FCM-S} &=& \mathcal{M}^{\mathcal{W}\mathcal{F}}_{FCM}-\mathcal{M}^{\mathcal{W}\mathcal{S}}_{FCM}\mathcal{R}^{\mathcal{E}\mathcal{S}}_{FCM}\mathcal{M}^{\mathcal{E}\mathcal{F}}_{FCM}, \\
\mathcal{M}^{\mathcal{W}\mathcal{T}}_{FCM-S} &=& \mathcal{M}^{\mathcal{W}\mathcal{T}}_{FCM}-\mathcal{M}^{\mathcal{W}\mathcal{S}}_{FCM}\mathcal{R}^{\mathcal{E}\mathcal{S}}_{FCM}\mathcal{M}^{\mathcal{E}\mathcal{T}}_{FCM}. 
\end{eqnarray}

In analysing fluctuating FCM, we will show that the random particle velocity correlations are given by the FCM mobility matrices, $\mathcal{M}_{FCM}$ and $\mathcal{M}_{FCM-S}$ and the resulting method is consistent with the fluctuation-dissipation theorem.

\section{The fluctuating force-coupling method}
Modifying FCM to include Brownian motion involves including a white-noise, fluctuating stress, $\mathbf{P}$, in the Stokes equations, so Eq. (\ref{eq:FCM1}) becomes
\begin{eqnarray}
-\bm{\nabla} p + \eta \nabla^2 \mathbf{u} &=& - \bm{\nabla}\cdot \mathbf{P} \nonumber\\
&&-\sum_n \left(\mathbf{F}_n \Delta_n(\mathbf{x})  + \frac{1}{2}\bm{\tau}_n \times \bm{\nabla} \Theta_n(\mathbf{x}) + \mathbf{S}_{n} \cdot \bm{\nabla}\Theta_n(\mathbf{x})\right) \nonumber \\
 \bm{\nabla}\cdot \mathbf{u} &=& 0.
\label{eq:RanVel1}
\end{eqnarray}
As introduced in \cite{LandauLifshitz,Fox1970}, the statistics for the fluctuating stress, in index notation, are given by
\begin{eqnarray}
	\left\langle P_{jl}\right\rangle&=&0\\
	\left\langle P_{jl}(\mathbf{x},t)P_{pq}(\mathbf{x}',t') \right\rangle&=&2k_BT\eta\left(\delta_{jp}\delta_{lq} + \delta_{jq}\delta_{lp}\right)\delta(\mathbf{x}-\mathbf{x}')\delta(t-t'). 
\label{eq:RanVel2}
\end{eqnarray}
Beyond this additional term, fluctuating FCM follows the same steps as the standard implementation of FCM.  After solving Eq. (\ref{eq:RanVel1}) for the fluid flow, the particle velocities, angular velocities, and local rates-of-strain are determined from Eqs. (\ref{eq:FCM3a}), (\ref{eq:FCM3b}), and (\ref{eq:FCM3c}).  Also, if the stresslets are included, we use the usual condition, $\mathbf{E}_n = \mathbf{0}$, to determine their entries.

\subsection{Particle velocity correlations} \label{sec:particle_cors}
While including fluctuations in FCM only involves forcing of the Stokes equations randomly, one must ensure that the resulting particle velocities and angular velocities do indeed satisfy the fluctuation-dissipation theorem.  In this section, we perform this analysis, taking $\mathbf{F}_n=\mathbf{0}$ and $\bm{\tau}_n = \mathbf{0}$ for each $n$.  

\subsubsection{Without particle stresslets}\label{sec:particle_cors_nostress}
If the particle stresslets are not included in the calculation, $\mathbf{S}_n = \mathbf{0}$ for each $n$, and the fluid is only driven by the fluctuating stress, $\mathbf{P}$.  Thus, Eq. (\ref{eq:RanVel1}) becomes
\begin{eqnarray}
-\bm{\nabla} p + \eta \nabla^2 \tilde\mathbf{u} + \bm{\nabla}\cdot \mathbf{P} &=& \mathbf{0} \\
 \bm{\nabla}\cdot \tilde\mathbf{u} &=& 0.
\label{eq:RanVel5}
\end{eqnarray}
We can show (see Appendix \ref{sec:flucstress}) the statistics of the resulting random fluid velocity, $\tilde{\mathbf{u}}$, will be given by
\begin{eqnarray}
\langle \tilde \mathbf{u}(\mathbf{x},t) \rangle &=& \mathbf{0} \\ 
\langle \tilde \mathbf{u}(\mathbf{x},t) \tilde\mathbf{u}^T(\mathbf{y},t')\rangle &=& 2k_BT \mathbf{G}(\mathbf{x}-\mathbf{y})\delta(t-t')
\label{eq:RanVel6}
\end{eqnarray}
where, again, $\mathbf{G}(\mathbf{x}-\mathbf{y})$ is the Stokeslet, see Eq. (\ref{eq:FCM6}).  From Eqs. (\ref{eq:FCM3a}) and (\ref{eq:FCM3b}), the particle velocities and angular velocities will be
\begin{eqnarray}
\tilde\mathbf{V}_n &=&\int \tilde\mathbf{u} \Delta_n(\mathbf{x})d^3\mathbf{x}, \nonumber \\
\tilde{\bm{\Omega}}_n &=& \frac{1}{2}\int \left[\bm{\nabla}\times\tilde{\mathbf{u}}\right] \Theta_n(\mathbf{x})d^3\mathbf{x}.
\label{eq:RanVel7}
\end{eqnarray}
Taking the ensemble average of these equations, we immediately see that $\langle \tilde{\mathbf{V}}_n\rangle = \mathbf{0}$ and $\langle \tilde{\bm{\Omega}}_n \rangle = \mathbf{0}$. 

We establish the velocity correlations between particles $n$ and $m$ by taking the ensemble average of the outer product of $\tilde{\mathbf{V}}_n$ and $\tilde{\mathbf{V}}_m$.  This will give us 
\begin{equation}
\langle \tilde{\mathbf{V}}_{n}(t) \tilde{\mathbf{V}}_{m}^T(t')\rangle = \int \int  \langle \tilde{\mathbf{u}}(\mathbf{x},t)\tilde{\mathbf{u}}^T(\mathbf{y},t')\rangle \Delta_m(\mathbf{y}) \Delta_n(\mathbf{x})d^3\mathbf{x} d^3\mathbf{y}.
\end{equation}
Using Eq. (\ref{eq:RanVel6}) for the correlations of the fluctuating flow field, this becomes
\begin{equation}
\langle \tilde{\mathbf{V}}_{n}(t) \tilde{\mathbf{V}}_{m}^T(t')\rangle = 2k_BT\delta(t-t')\int \int  \mathbf{G}(\mathbf{x} - \mathbf{y}) \Delta_m(\mathbf{y}) \Delta_n(\mathbf{x})d^3\mathbf{x} d^3\mathbf{y}.
\end{equation}
We recognise the double integral as the entries of $\mathcal{M}^{\mathcal{V}\mathcal{F}}_{FCM}$, the submatrix of the FCM mobility matrix $\mathcal{M}_{FCM}$, that relate the velocities of particle $n$ and the forces on particle $m$, see Eq. (\ref{eq:FCM8}).  Thus, taking into account all particle pairs, we will have
\begin{equation} 
\langle \tilde{\mathcal{V}}(t) \tilde{\mathcal{V}}^T(t')\rangle = 2k_BT \mathcal{M}^{\mathcal{VF}}_{FCM}\delta(t-t')
\label{eq:RanVel8}
\end{equation}
By a similar analysis, see Appendix \ref{sec:FCMcorr_app}, the angular-angular and linear-angular velocity correlations are shown to be 
\begin{eqnarray}
\langle \tilde{\mathcal{W}}(t) \tilde{\mathcal{W}}^T(t') \rangle &=& 2k_BT\mathcal{M}^{\mathcal{WT}}_{FCM}\delta(t-t') \\
\langle \tilde{\mathcal{V}}(t) \tilde{\mathcal{W}}^T(t') \rangle &=& 2k_BT\mathcal{M}^{\mathcal{VT}}_{FCM}\delta(t-t'),
\end{eqnarray}
and, consequently,
\begin{eqnarray}
\left\langle \left[\begin{array}{c}
\tilde{\mathcal{V}}(t) \\
\tilde{\mathcal{W}}(t) \\
\end{array}\right] 
\left[\begin{array}{cc}
\tilde{\mathcal{V}}^T(t') & \tilde{\mathcal{W}}^T(t')
\end{array}\right]
\right\rangle &=& 2k_BT \mathcal{M}_{FCM}\delta(t-t')
\end{eqnarray}

\subsubsection{With particle stresslets}\label{sec:particle_cors_stress}
When we include the stresslets in fluctuating FCM, the resulting fluid velocity may be expressed as
\begin{equation}
\mathbf{u}(\mathbf{x},t) = \tilde{\mathbf{u}}(\mathbf{x},t)  + \sum_{m}\int \mathbf{G}(\mathbf{x} - \mathbf{y})\cdot \tilde{\mathbf{S}}_m \cdot \bm{\nabla} \Theta_m(\mathbf{y}) d^3\mathbf{y}.
\end{equation}
We must first determine the unknown stresslets by inserting this expression for the fluid velocity into Eq. (\ref{eq:FCM3c}) and setting the resulting local rate-of-strain equal to zero.  This gives us a linear system, and after solving it, we find the stresslets, in matrix representation, to be
\begin{equation}
\tilde{\mathcal{S}} = -\mathcal{R}^{\mathcal{ES}}_{FCM}\tilde{\mathcal{E}} \label{eq:stresslets}
\end{equation}
where the $5N \times 1$ vector $\tilde{\mathcal{E}}$ holds the independent components of the random local rate-of-strain,
\begin{equation}
\tilde{\mathbf{E}}_{n} = \frac{1}{2}\int \left[\bm{\nabla}\tilde\mathbf{u} + (\bm{\nabla}\tilde\mathbf{u})^T\right]\Theta_n(\mathbf{x})d^3\mathbf{x}
\end{equation}
for all of the particles.  As demonstrated in Appendix \ref{sec:FCMcorr_app}, $\tilde\mathcal{E}$ has the following correlations with $\tilde\mathcal{V}$, $\tilde\mathcal{W}$, and itself
\begin{eqnarray}
\langle \tilde{\mathcal{V}}(t)\tilde{\mathcal{E}}^T(t')\rangle&=&-2k_BT \mathcal{M}^{\mathcal{VS}}_{FCM}\delta(t-t'), \nonumber \\
\langle \tilde{\mathcal{W}}(t)\tilde{\mathcal{E}}^T(t')\rangle&=&-2k_BT \mathcal{M}^{\mathcal{WS}}_{FCM}\delta(t-t'), \nonumber \\
\langle \tilde{\mathcal{E}}(t) \tilde{\mathcal{E}}^T(t')\rangle&=&-2k_BT \mathcal{M}^{\mathcal{ES}}_{FCM}\delta(t-t'),
\label{eq:Ecor}
\end{eqnarray}
while from Eq. (\ref{eq:RanVel6}), we see immediately that $\langle \tilde{\mathcal{E}} \rangle = \mathbf{0}$.

With the stresslets known, the velocities and angular velocities are given by 
\begin{eqnarray}
\tilde{\mathcal{V}}_S&=&\tilde{\mathcal{V}} + \mathcal{M}^{\mathcal{VS}}_{FCM}\tilde\mathcal{S} \\
\tilde{\mathcal{W}}_S&=&\tilde{\mathcal{W}} + \mathcal{M}^{\mathcal{WS}}_{FCM}\tilde\mathcal{S}
\end{eqnarray}
and we can now determine the particle velocity correlations when the stresslets are included in fluctuating FCM.  Taking the ensemble average of the outer product of $\tilde{\mathcal{V}}_S$ with itself gives
\begin{eqnarray}
\langle \tilde{\mathcal{V}}_S(t) \tilde{\mathcal{V}}_S^T(t')\rangle &=&\langle \tilde{\mathcal{V}}(t) \tilde{\mathcal{V}}^T(t')\rangle + \langle \tilde{\mathcal{V}}(t)  (\mathcal{M}^{\mathcal{VS}}_{FCM}\tilde\mathcal{S}(t'))^T \rangle\nonumber \\
&& +\langle \mathcal{M}^{\mathcal{VS}}_{FCM}\tilde\mathcal{S}(t) \tilde{\mathcal{V}}^T(t')\rangle + \langle \mathcal{M}^{\mathcal{VS}}_{FCM}\tilde\mathcal{S}(t) (\mathcal{M}^{\mathcal{VS}}_{FCM}\tilde\mathcal{S}(t'))^T\rangle. \nonumber\\
\end{eqnarray}
We then substitute Eq. (\ref{eq:stresslets}) for $\tilde\mathcal{S}$ and rearrange terms to find
\begin{eqnarray}
\langle \tilde{\mathcal{V}}_S(t) \tilde{\mathcal{V}}_S^T(t')\rangle &=&\langle \tilde{\mathcal{V}}(t) \tilde{\mathcal{V}}^T(t')\rangle - \langle \tilde{\mathcal{V}}(t) \tilde{\mathcal{E}}^T(t')\rangle(\mathcal{R}^{\mathcal{ES}}_{FCM})^T(\mathcal{M}^{\mathcal{VS}}_{FCM})^T \nonumber\\
&& - \mathcal{M}^{\mathcal{VS}}_{FCM}\mathcal{R}^{\mathcal{ES}}_{FCM} \langle \tilde{\mathcal{E}}(t) \tilde{\mathcal{V}}^T(t')\rangle \nonumber\\
&&+ \mathcal{M}^{\mathcal{VS}}_{FCM}\mathcal{R}^{\mathcal{ES}}_{FCM}\langle \tilde{\mathcal{E}}(t) \tilde{\mathcal{E}}^T(t')\rangle (\mathcal{R}^{\mathcal{ES}}_{FCM})^T(\mathcal{M}^{\mathcal{VS}}_{FCM})^T.
\label{eq:VScor1}
\end{eqnarray}
From the velocity and rate of strain correlations, Eqs. (\ref{eq:RanVel8}) and (\ref{eq:Ecor}) respectively, and the fact that $\langle \tilde{\mathcal{E}} \tilde{\mathcal{V}}^T\rangle = (\langle \tilde{\mathcal{V}} \tilde{\mathcal{E}}^T\rangle)^T$, Eq. (\ref{eq:VScor1}) becomes
\begin{eqnarray}
\frac{\langle \tilde{\mathcal{V}}_S(t) \tilde{\mathcal{V}}_S^T(t') \rangle}{2k_BT} &=\delta(t-t')&\Bigg[\mathcal{M}^{\mathcal{VF}}_{FCM} + \mathcal{M}^{\mathcal{VS}}_{FCM}(\mathcal{R}^{\mathcal{ES}}_{FCM})^T(\mathcal{M}^{\mathcal{VS}}_{FCM})^T \nonumber \\
&& +\mathcal{M}^{\mathcal{VS}}_{FCM}\mathcal{R}^{\mathcal{ES}}_{FCM} (\mathcal{M}^{\mathcal{VS}}_{FCM})^T \nonumber\\
&& -\mathcal{M}^{\mathcal{VS}}_{FCM}\mathcal{R}^{\mathcal{ES}}_{FCM}\mathcal{M}^{\mathcal{ES}}_{FCM}(\mathcal{R}^{\mathcal{ES}}_{FCM})^T(\mathcal{M}^{\mathcal{VS}}_{FCM})^T\Bigg].\nonumber\\
\end{eqnarray}
Finally, using the following properties of the FCM matrices, $(\mathcal{M}^{\mathcal{VS}}_{FCM})^T = -\mathcal{M}^{\mathcal{EF}}_{FCM}$, $(\mathcal{R}^{\mathcal{ES}}_{FCM})^T=\mathcal{R}^{\mathcal{ES}}_{FCM}$, and $\mathcal{R}^{\mathcal{ES}}_{FCM}=(\mathcal{M}^{\mathcal{ES}}_{FCM})^{-1}$, we arrive at
\begin{eqnarray}
\langle \tilde{\mathcal{V}}_S(t) \tilde{\mathcal{V}}_S^T(t') \rangle/(2k_BT)&=&\delta(t-t')\left[\mathcal{M}^{\mathcal{VF}}_{FCM} - \mathcal{M}^{\mathcal{VS}}_{FCM}\mathcal{R}^{\mathcal{ES}}_{FCM}\mathcal{M}^{\mathcal{EF}}_{FCM}\right] \nonumber \\
&=&\delta(t-t') \mathcal{M}^{\mathcal{VF}}_{FCM-S}. 
\end{eqnarray}
Though not shown, repeating the same calculation for $\langle \mathcal{W}_S(t) \mathcal{W}^T_S(t')\rangle$ and $\langle \mathcal{V}_S(t) \mathcal{W}_S^T(t')\rangle$, one finds that 
\begin{eqnarray}
\langle \mathcal{W}_S(t) \mathcal{W}^T_S(t')\rangle/(2k_BT)&=&\delta(t-t')\mathcal{M}^{\mathcal{WT}}_{FCM-S} \\
\langle \mathcal{V}_S(t) \mathcal{W}_S^T(t')\rangle/(2k_BT)&=&\delta(t-t')\mathcal{M}^{\mathcal{VT}}_{FCM-S}.
\end{eqnarray}
Putting all of these results together, we see that
\begin{eqnarray}
\left\langle \left[\begin{array}{c}
\tilde{\mathcal{V}}_S(t) \\
\tilde{\mathcal{W}}_S(t) \\
\end{array}\right] 
\left[\begin{array}{cc}
\tilde{\mathcal{V}}^T_S(t') & \tilde{\mathcal{W}}^T_S(t')
\end{array}\right]
\right\rangle &=& 2k_BT \mathcal{M}_{FCM-S}\delta(t-t'),
\end{eqnarray}
satisfying the fluctuation-dissipation theorem.

\section{Discretisation of fluctuating FCM}

In our simulations, we use a Fourier spectral method to solve the Stokes equations, Eq. (\ref{eq:RanVel1}), on a triply periodic domain.  Each side of the domain has length $L = 2\pi$ and we use $M$ grid points in each direction, giving a total number of $N_g = M^3$ points.  This sets the grid spacing to be $h = 2\pi/M$ and the grid points as $x_\alpha = \alpha h$ for $\alpha = 0, \dots, M-1$.  The corresponding wave numbers are 
\begin{equation}
k_\alpha = 
\Bigg \{ \begin{array}{ll}
\alpha, & 0 \leq \alpha \leq M/2 \\
\alpha - M, & M/2 + 1 \leq \alpha \leq M-1.  
\end{array}
\label{eq:wavenum}
\end{equation}

While we can utilise many of the numerical techniques typically employed with FCM, see for example \cite{Maxey2001,Lomholt2003,Dance2003,Yeo2010}, the statistics for the fluctuating stress, Eq. (\ref{eq:RanVel2}), must be handled appropriately in the discretised system.  Here, we follow other numerical methods where fluctuating stresses are considered, especially DLM \cite{Sharma2004} and the immersed boundary method \cite{Atzberger2007}.   At each grid point, the fluctuating stress is an independent Gaussian random variable with the following statistics 
\begin{eqnarray}
	\langle P_{ij}(x_\alpha, x_\beta, x_\gamma)\rangle&=&0\\
	\left\langle P_{ij}(x_\alpha, x_\beta, x_\gamma)P_{pq}(x_\alpha, x_\beta ,x_\gamma) \right\rangle&=&\frac{2k_BT\eta}{h^3 \Delta t}\left(\delta_{ip}\delta_{jq} + \delta_{iq}\delta_{jp}\right).
\label{eq:FCMdis1}	
\end{eqnarray}
where $\Delta t$ is the timestep.  The computational cost associated with this step of the calculation is $\mathcal{O}(N_g)$.  With the discrete Fourier transform (DFT) and the inverse DFT defined as
\begin{eqnarray}
\hat{g}(k_\alpha, k_\beta, k_\gamma)&=&\sum_{\zeta} \sum_{\xi} \sum_{\lambda} g(x_\zeta, x_\xi, x_\lambda) e^{-i(k_\alpha x_\zeta + k_\beta x_\xi + k_\gamma x_\lambda)}, \\
g(x_\alpha, x_\beta, x_\gamma)&=&\frac{1}{M^3}\sum_{\zeta} \sum_{\xi} \sum_{\lambda} \hat{g}(k_\zeta, k_\xi, k_\lambda) e^{i(k_\zeta x_\alpha  + k_\xi x_\beta + k_\lambda x_\gamma)},
\end{eqnarray}
Eq. (\ref{eq:FCMdis1}) will be
\begin{eqnarray}
	\left\langle \hat{P}_{ij}(k_\alpha, k_\beta, k_\gamma)\right\rangle&=&0\\
	\left\langle \hat{P}_{ij}(k_\alpha, k_\beta, k_\gamma)P_{pq}(-k_\alpha, -k_\beta, -k_\gamma) \right\rangle&=&\frac{2k_BT\eta M^3}{h^3 \Delta t}\left(\delta_{ip}\delta_{jq} + \delta_{iq}\delta_{jp}\right)
\end{eqnarray}
in the discrete Fourier Space.  

After generating the Gaussian random variables for the fluctuating stress, we then evaluate the FCM force distribution, 
\begin{equation}
\mathbf{f}_{FCM}(\mathbf{x}) = \sum_n \mathbf{F}_n \Delta_n(\mathbf{x})  - \frac{1}{2}\bm{\tau}_n \times \bm{\nabla} \Theta_n(\mathbf{x}) + \mathbf{S}_{n} \cdot \bm{\nabla}\Theta_n(\mathbf{x}),
\end{equation}
at the grid points.  Since we may assume that for the rapidly decaying Gaussian functions $\Delta_n(\mathbf{x}) = 0$ and $ \Theta_n(\mathbf{x}) = 0$ for $|\mathbf{x} - \mathbf{Y}_n| > 3a$, \cite{Yeo2010} this stage of the calculation can be done in $\mathcal{O}(N)$ operations.  We then take the DFT of the total force distribution, an $\mathcal{O}(N_g \textrm{log} N_g)$ calculation using FFTs, and compute the DFT of the incompressible velocity field,
\begin{eqnarray}
\hat{\mathbf{u}}(k_{\alpha}, k_{\beta}, k_{\gamma})&=& \frac{1}{\eta |\mathbf{k}|^2} \left(\mathbf{I} - \frac{\mathbf{k}\mathbf{k}}{|\mathbf{k}|^2} \right)\left(i\mathbf{k} \cdot \hat{\mathbf{P}}(k_{\alpha}, k_{\beta}, k_{\gamma}) + \hat{\mathbf{f}}_{FCM}(k_{\alpha}, k_{\beta}, k_{\gamma})\right), \nonumber \\
\label{eq:DFTu}
\end{eqnarray}
where $\mathbf{k} = [\begin{array}{ccc}k_{\alpha} & k_{\beta} & k_{\gamma}\end{array}]^{T}$.  The fluid velocity at the grid points, $ \mathbf{u}(x_{\alpha}, x_{\beta}, x_{\gamma})$, is found by taking the inverse DFT of Eq. (\ref{eq:DFTu}) in $\mathcal{O}(N_g \textrm{log} N_g)$ operations.  The velocity, angular velocity, and local rate-of-strain for each particle are then computed by applying the spectrally accurate trapezoidal rule to Eqs. (\ref{eq:FCM3a}) -- (\ref{eq:FCM3c}).  Again, with the rapid decay of the Gaussian envelopes, we may set $\Delta_n(\mathbf{x}) = 0$ and $ \Theta_n(\mathbf{x}) = 0$ for $|\mathbf{x} - \mathbf{Y}_n| > 3a$, so the volume averaging incurs an $\mathcal{O}(N)$ computational cost.  In our simulations, we fix $\sigma_{\Theta}/h = 1.5$ and $\sigma_{\Delta}/h = 1.86$.  Therefore, if we keep the volume fraction constant while increasing $N$, $N_g$ will increase linearly with $N$ and the overall computational cost will be $\mathcal{O}(N \textrm{log}N)$.  
For the simulations where the particle stresslets are included, we employ the conjugate gradient scheme detailed in \cite{Yeo2010} to obtain $\mathbf{S}_n$ for each $n$.  Each iteration requires $\mathcal{O}(N_g \textrm{log}N_g)$ operations.

\section{Convergence}
\begin{figure}
	\begin{center}
	\subfigure[\label{fig:ModeError}]{\includegraphics[width=16.25pc]{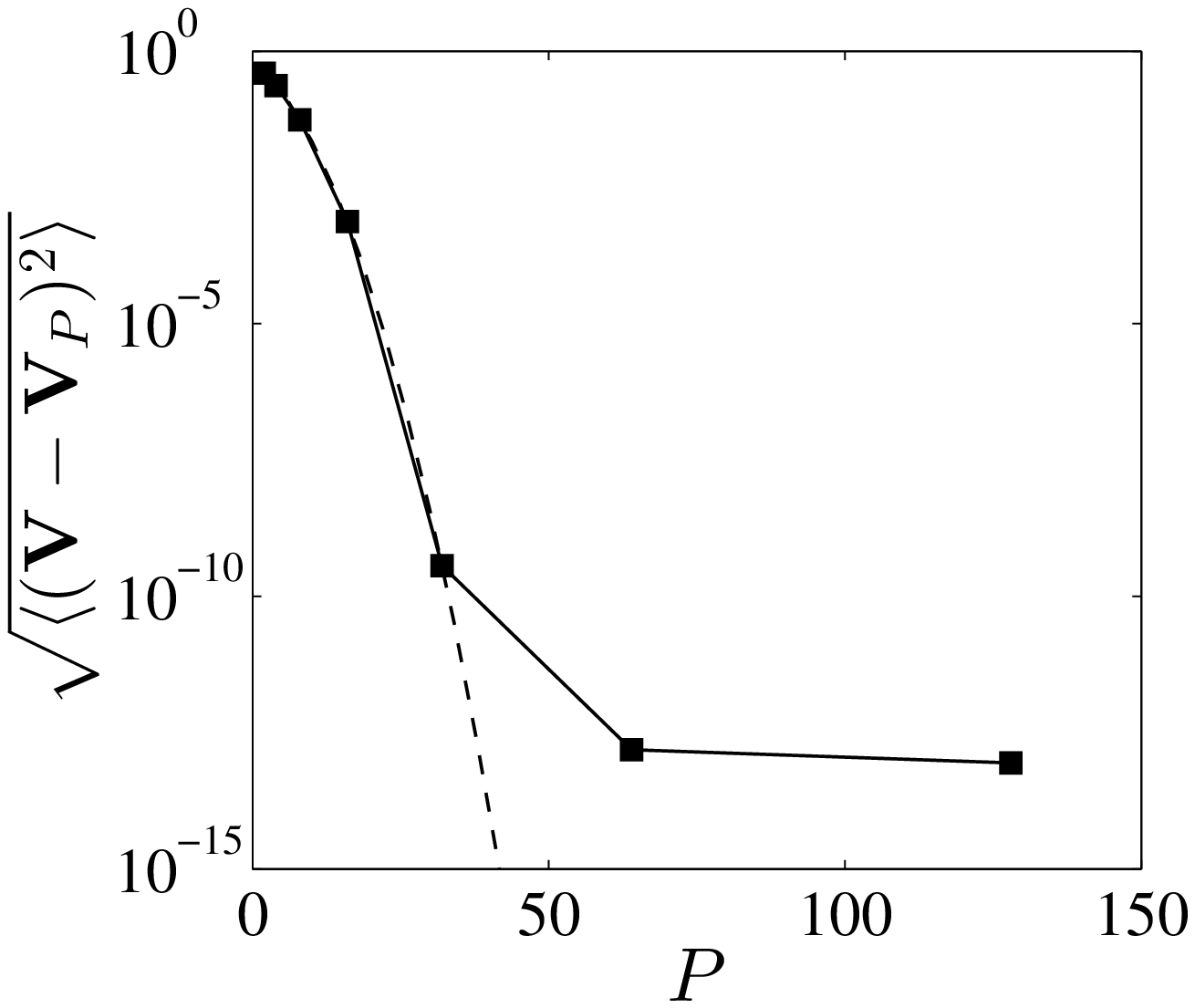}}
	\subfigure[\label{fig:VarError}]{\includegraphics[width=16.25pc]{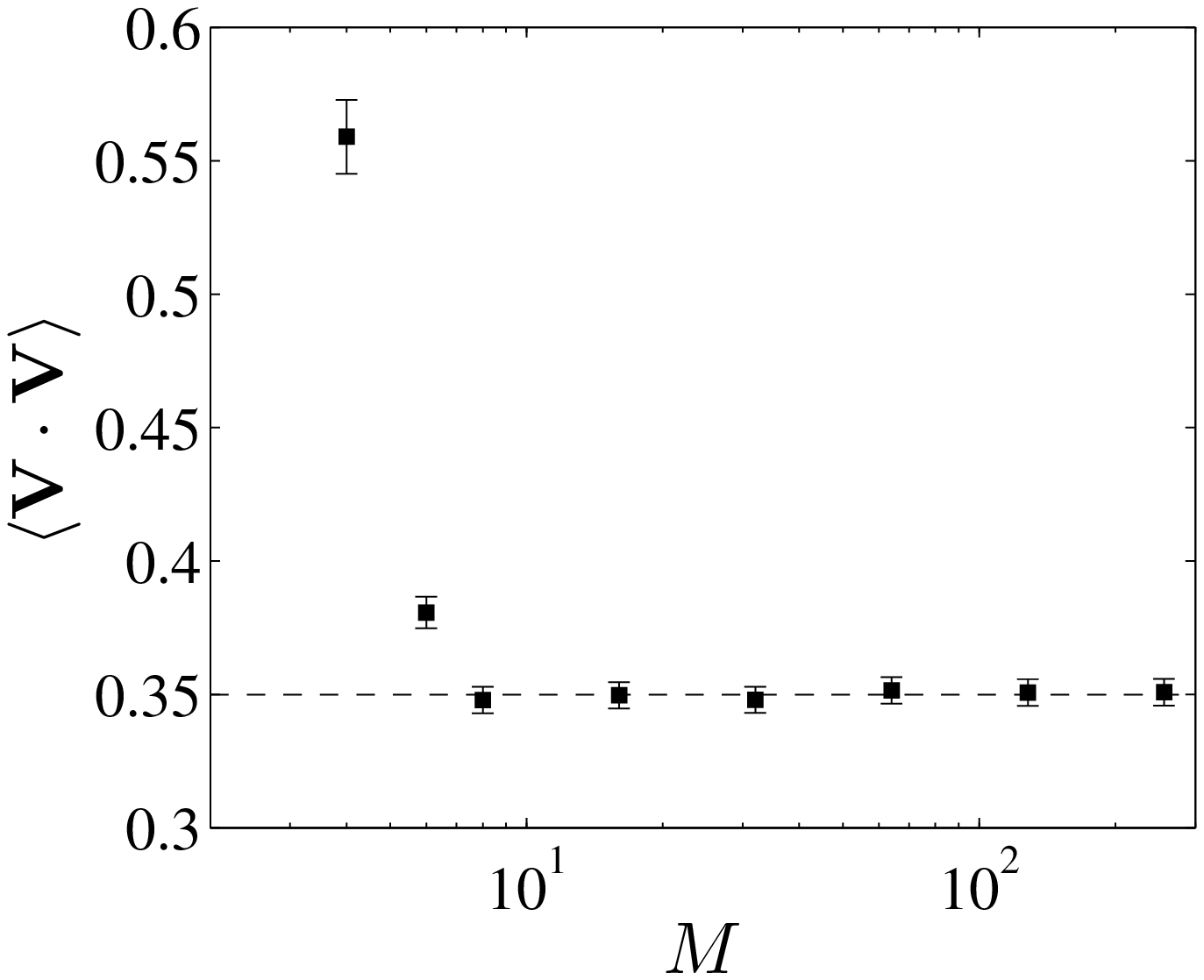}}
	\caption{(a) Root mean squared error in the particle velocity as a function of the number of random modes, $P$, with $M = 256$.  The symbols show the computed error while the dashed line gives the error estimate based on a truncated Fourier series, Eq. (\ref{eq:fourerrest}). (b) The computed value of the velocity correlations as a function of $M$.  The dashed line indicates the exact value.}
	\label{fig:convergence}
	\end{center}
\end{figure}

In order to obtain the correct random particle motion, it is important to include a sufficient number of modes for the random flow.  We can analyse the dependence of $\mathbf{V}$ on the number of these modes by considering a single particle in a periodic domain.  If $\mathbf{F} = \mathbf{0}$ and $\bm{\tau} = \mathbf{0}$, the particle velocity can be written as
\begin{equation}
\mathbf{V} =\frac{1}{(2\pi)^3}\sum_{\zeta=-\infty}^{\infty} \sum_{\xi=-\infty}^{\infty} \sum_{\lambda=-\infty}^{\infty} \hat{\Delta}(\mathbf{k}) \hat{\tilde\mathbf{u}}(\mathbf{k})
\end{equation}
where $\mathbf{k} = [k_\zeta, k_\xi, k_\lambda]^T$ and the Fourier coefficients $\hat{\Delta}(\mathbf{k})$ and $\hat{\tilde\mathbf{u}}(\mathbf{k})$ are given by
\begin{eqnarray}
\hat{\Delta}(\mathbf{k}) &=& \int \Delta(\mathbf{x}) e^{-i\mathbf{k}\cdot \mathbf{x}}d^3\mathbf{x}\\
\hat{\tilde{\mathbf{u}}}(\mathbf{k}) &=& \int \tilde{\mathbf{u}}(\mathbf{x})  e^{-i\mathbf{k}\cdot \mathbf{x}}d^3\mathbf{x}.
\end{eqnarray}
with the integrals being performed over a $2\pi^3$ domain.  If we limit the number of random flow modes to the lowest $P+1$ modes in each direction, the particle velocity will be given by the truncated series
\begin{equation}
\mathbf{V}_{P} =\frac{1}{(2\pi)^3}\sum_{|\zeta|\leq P/2} \sum_{|\xi| \leq P/2} \sum_{|\lambda| \leq P/2} \hat{\Delta}(\mathbf{k})\hat{\tilde\mathbf{u}}(\mathbf{k}),
\end{equation}
and, we have that 
\begin{equation}
\mathbf{V} - \mathbf{V}_P =\frac{1}{(2\pi)^3}\sum_{|\zeta| > P/2} \sum_{|\xi| > P/2} \sum_{|\lambda| > P/2} \hat{\Delta}(\mathbf{k}) \hat{\tilde\mathbf{u}}(\mathbf{k}).
\end{equation}
Taking the ensemble average of $(\mathbf{V} - \mathbf{V}_{P})^2$ and using the correlations for $\hat{\tilde\mathbf{u}}(\mathbf{k})$ gives
\begin{equation}
\langle (\mathbf{V} - \mathbf{V}_{P})^2 \rangle=\frac{2k_BT}{(2\pi)^3}\sum_{|\zeta| > P/2} \sum_{|\xi| > P/2} \sum_{|\lambda| > P/2}\left[\hat{\Delta}(\mathbf{k})\right]^2 \textrm{trace}(\hat{\mathbf{G}}(\mathbf{k})).
\end{equation}
As $\textrm{trace}(\hat{\mathbf{G}}(\mathbf{k})) = 2/(\eta k^2)$, 
\begin{equation}
\langle (\mathbf{V} - \mathbf{V}_{P})^2 \rangle=\frac{4k_BT}{\eta(2\pi)^3}\sum_{|\zeta| > P/2} \sum_{|\xi| > P/2} \sum_{|\lambda| > P/2} \left[\hat{\Delta}(\mathbf{k})\right]^2/k^2.
\end{equation}
For highly localised Gaussian distributions where, $\sigma_\Delta \ll \pi$, we may approximate $\hat{\Delta}(\mathbf{k}) \approx e^{-k^2\sigma_\Delta^2/2}$ to obtain the estimate
\begin{eqnarray}
\langle (\mathbf{V} - \mathbf{V}_{M})^2 \rangle_{EST} &=& \frac{4k_BT}{\eta(2\pi)^3}\sum_{|\zeta| > P/2} \sum_{|\xi| > P/2} \sum_{|\lambda| > P/2} e^{-k^2 \sigma_{\Delta}^2}/k^2. \label{eq:fourerrest}
\end{eqnarray}
We have performed a series of computations where for each realisation of the random flow, we computed $\mathbf{V}_{P}$ for different values of $P$.  For these computations, $M=256$ and $\sigma_\Delta/\pi = 0.12$.  For these values, we also have that, $\sigma_\Delta/h = 14.89$ which is sufficient to reduce any error in $\hat\Delta(\mathbf{k})$ from the DFT to machine precision.  Thus, the error we observe should come purely from the neglected random flow modes.  The RMS error for these computations, as well as the values given by our estimate, Eq. (\ref{eq:fourerrest}), are shown in Fig. \ref{fig:ModeError}.  We find that the RMS error decays rapidly as we increase $P$, with the error for $P=32$ being $3.6\times10^{-10}$.  

In simulations, however, one has $P = M$ and the resolution of the Gaussian envelope is tied to the number of modes for the random flow.  To better understand this joint dependence, we computed the mean squared particle velocity using for different values of $M$.  For each case, $10^4$ realisations of the flow and random particle positions are used for the ensemble averaging.  The results from these computations are shown in Fig. \ref{fig:VarError}.  We see that even when varying $M$ itself, we still recover an accurate value of the mean square velocity for values of $M$ as low as $M = 8$.  This corresponds to the value $\sigma_{\Delta} = 0.47h$, which, based on our experience, is too low to accurately resolve particle motion when forces are also present.  Thus, as is typically done with deterministic simulations FCM, we set $\sigma_{\Delta} = 1.86h$ and $\sigma_{\Theta} = 1.5h$.  This value provides sufficient accuracy while keeping the computational costs low.  The number of independent particles for a given volume fraction is, therefore, set by $M.$  For our simulations, we take $M=64$, except for the periodic array of spheres computations where we have $M=32$ and vary $\sigma_\Delta$ to obtain the desired volume fraction. 

\section{Time integration and Brownian drift}

To simulate the dynamics of a Brownian suspension using fluctuating FCM, one must also account for the Brownian drift term, $k_BT\nabla_{\mathcal{Y}}\cdot \mathcal{M}^{\mathcal{VF}}$ in Eq. (\ref{eq:PM1}).  For the case where the particle stresslets are ignored, $\mathbf{S}_n = \mathbf{0}$, the entries of the corresponding FCM mobility matrix $\mathcal{M}^{\mathcal{VF}}_{FCM}$ come directly from solutions to the Stokes equations.  Since these solutions satisfy the divergence free condition and are translationally invariant for periodic boundary conditions, the Brownian Drift term will be identically zero. To update the particle positions, we integrate the equations of motion using the forward Euler-Maruyama scheme,
\begin{equation}
\mathcal{Y}_{k+1} = \mathcal{Y}_{k} + \left(\mathcal{V}_k + \tilde \mathcal{V}_k\right)\Delta t.
\end{equation}
in the manner typically used for Brownian Dynamics simulations, as described by Ermak and McCammon \cite{Ermak1978}.

\begin{figure}
	\begin{center}
	\includegraphics[width=16.25pc]{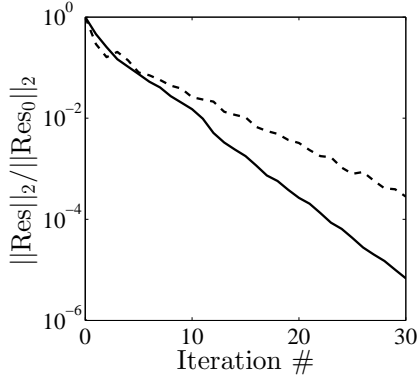}
	\caption{$L_2$ norm of the residual versus iteration number.  The dashed line corresponds to no preconditioning, while the solid line indicates the residual when the preconditioner was used.}
	\label{fig:cg_iter}
	\end{center}
\end{figure}

For simulations where the stresslets are included, the Brownian drift term is no longer zero.  We can, however, avoid computing it directly by using the first-order midpoint integration scheme introduced by Fixman \cite{Fixman1978,Grassia1995}.  Specifically, the particle positions are updated using the scheme
\begin{eqnarray}
\mathcal{Y}_{k+1/2} &=& \mathcal{Y}_{k} + \frac{\Delta t}{2}\left(\mathcal{V}_k + \tilde \mathcal{V}_k\right)\\
\mathcal{Y}_{k+1} &=&  \mathcal{Y}_{k} + \Delta t\left(\mathcal{V}_{k+1/2} + \tilde \mathcal{V}_{k+1/2}\right)
\end{eqnarray}
where
\begin{eqnarray}
\mathcal{V}_{k+1/2}&=& \mathcal{M}^{\mathcal{VF}}_{FCM-S; k+1/2}\mathcal{F}_k + \mathcal{M}^{\mathcal{VT}}_{FCM-S; k+1/2}\mathcal{T}_{k}\\
\tilde{\mathcal{V}}_{k+1/2}&=&\mathcal{M}^{\mathcal{VF}}_{FCM-S; k+1/2}\tilde{\mathcal{F}}_k + \mathcal{M}^{\mathcal{VT}}_{FCM-S;k+1/2}\tilde{\mathcal{T}}_{k}
\end{eqnarray}
and $\mathcal{M}^{\mathcal{VF}}_{FCM-S; k+1/2}$ and $\mathcal{M}^{\mathcal{VT}}_{FCM-S; k+1/2}$ are the mobility matrices based on the particle positions $\mathcal{Y}_{k+1/2}$.  While this integration scheme circumvents the direct calculation of the Brownian drift term, it utilises the random forces $\tilde \mathcal{F}_k$ and torques $\tilde \mathcal{T}_k$ at time $t_k$.  We can, however, find $\tilde \mathcal{F}_k$ and $\tilde \mathcal{T}_k$ from $\tilde \mathcal{V}_k$ and $\tilde \mathcal{W}_k$ by solving the linear system,
\begin{equation}
\left[\begin{array}{c}
\tilde{\mathcal{V}}_k \\
\tilde{\mathcal{W}}_k \\
-\tilde{\mathcal{E}}_k
\end{array}\right]=
\left[\begin{array}{ccc}
\mathcal{M}^{\mathcal{V}\mathcal{F}}_{FCM;k} & \mathcal{M}^{\mathcal{V}\mathcal{T}}_{FCM;k}   & \mathcal{M}^{\mathcal{V}\mathcal{S}}_{FCM;k}  \\
\mathcal{M}^{\mathcal{W}\mathcal{F}}_{FCM;k}  & \mathcal{M}^{\mathcal{W}\mathcal{T}}_{FCM;k}  & \mathcal{M}^{\mathcal{W}\mathcal{S}}_{FCM;k}  \\
-\mathcal{M}^{\mathcal{E}\mathcal{F}}_{FCM;k}  & -\mathcal{M}^{\mathcal{E}\mathcal{T}}_{FCM;k}  & -\mathcal{M}^{\mathcal{E}\mathcal{S}}_{FCM;k} 
\end{array}\right]
\left[\begin{array}{c}
\tilde{\mathcal{F}}_k \\
\tilde{\mathcal{T}}_k \\
\tilde{\mathcal{S}}_k
\end{array}\right].
\label{eq:BD}
\end{equation}
This relationship comes directly from Eq. (\ref{eq:FCM5}), the grand mobility matrix for FCM \cite{Yeo2010}, with the last line multiplied by negative one.  This transforms the linear system into one that is symmetric positive definite \cite{Yeo2010} and allows us to determine $\tilde \mathcal{F}_k$ and $\tilde \mathcal{T}_k$ efficiently using the conjugate gradient method.  Since the diagonal elements of $\mathcal{M}^{\mathcal{V}\mathcal{F}} \sim a^{-1}$, while $\mathcal{M}^{\mathcal{W}\mathcal{T}} \sim a^{-3}$ and $\mathcal{M}^{\mathcal{E}\mathcal{S}} \sim a^{-3}$, we expect the condition number to scale like $\kappa(\mathcal{M}) \sim a^2$ for dilute suspensions.  Thus, it can also be useful to use a preconditioner because with $\sigma_\Theta = 1.5h$, $a = 3.3h < 1$.  For the preconditioner, we utilise a diagonal matrix based on the mobility coefficients for a single particle in a periodic domain.  Specifically, we have
\begin{eqnarray}
\mathcal{M}^{\mathcal{V}\mathcal{F}}_{PRE} &=& \frac{\gamma}{6\pi \eta a} \mathcal{I} \\
\mathcal{M}^{\mathcal{W}\mathcal{T}}_{PRE} &=& \frac{1}{8\pi \eta a^3} \mathcal{I} \\
\mathcal{M}^{\mathcal{E}\mathcal{S}}_{PRE} &=& \frac{3}{20\pi \eta a^3} \mathcal{I} 
\end{eqnarray}
where in each case $\mathcal{I}$ is the identity matrix of the appropriate size and the coefficient $\gamma$, as described in the next section, is the modification of the Stokes drag law due to domain periodicity.  Fig. \ref{fig:cg_iter} shows the $L_2$ norm of the residual versus the number of conjugate gradient iterations with and without the preconditioner.  For both cases, we used identical realisations of the fluctuating stress field and the same random positions of $N = 183$ particles, corresponding to a volume fraction of $\phi = 0.10$.   The preconditioner provides faster convergence, particularly when the residual is less than 1\%.  It is worth noting that instead of using a preconditioner, one could alternatively set $a=1$ and rescale the domain length, $L$, and the wave numbers, Eq. (\ref{eq:wavenum}).  This would remove the dependence of the mobility matrix condition number on the grid spacing.

\section{Fluctuating FCM simulations}
We perform a series of simulations to confirm the analytical results presented above and demonstrate the effectiveness of fluctuating FCM.  We compute both the short-time and long-time diffusion coefficients for suspensions of interacting particles and determine the equilibrium concentration profiles for Brownian suspensions subject to an external potential.  For these simulations, we compare the fluctuating FCM results with those found analytically, or with numerical results from studies that employed Brownian or Stokesian dynamics.   In addition, we show how fluctuating FCM can be used to explore the dynamics of suspensions in periodic cellular flow fields, highlighting the role of hydrodynamic interactions and how they affect particle diffusion.

\subsection{Short-time self-diffusion coefficient}
\begin{figure}
	\begin{center}
	\subfigure[\label{fig:lattice}]{\includegraphics[width=16.25pc]{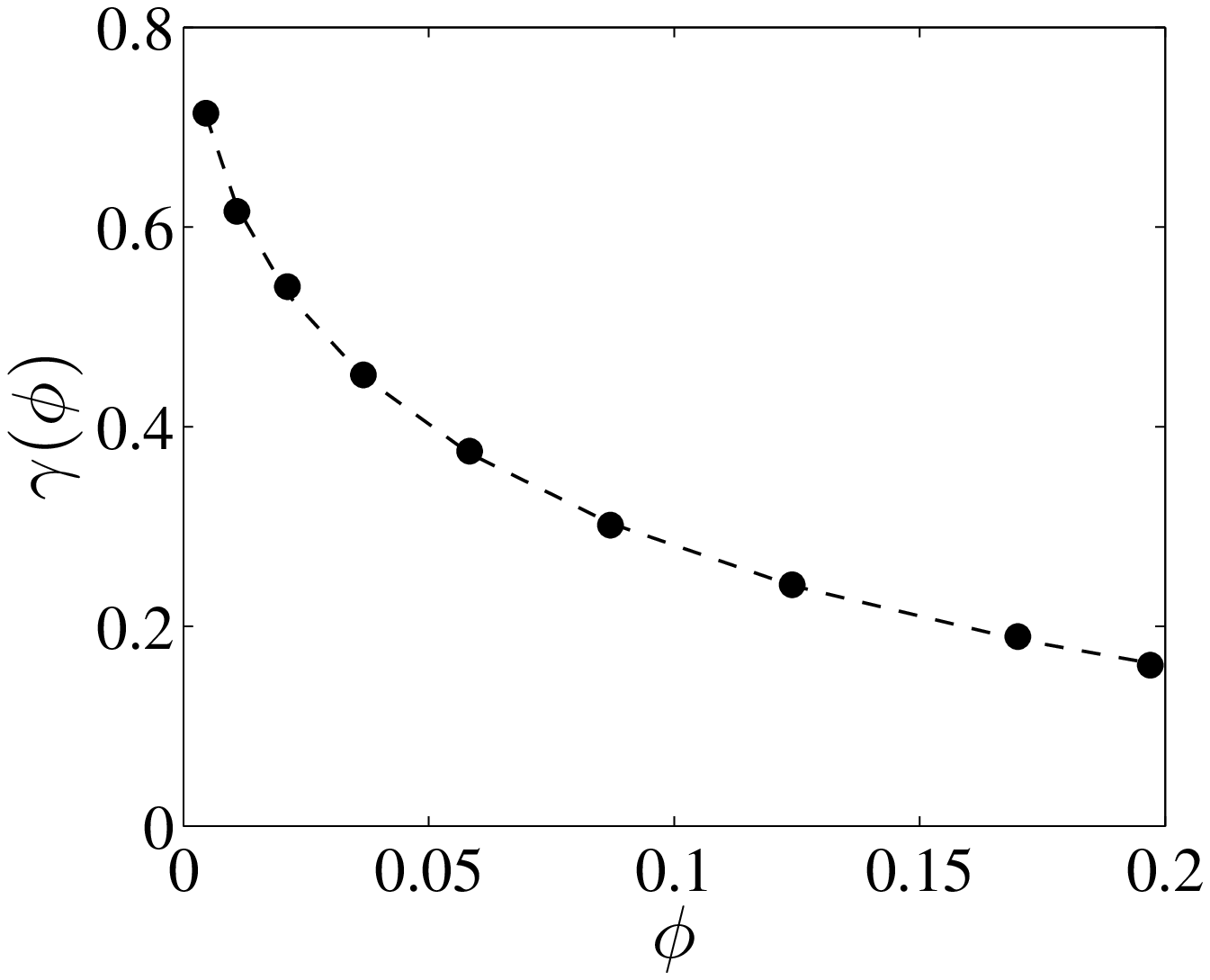}}
	\subfigure[\label{fig:std}]{\includegraphics[width=16.25pc]{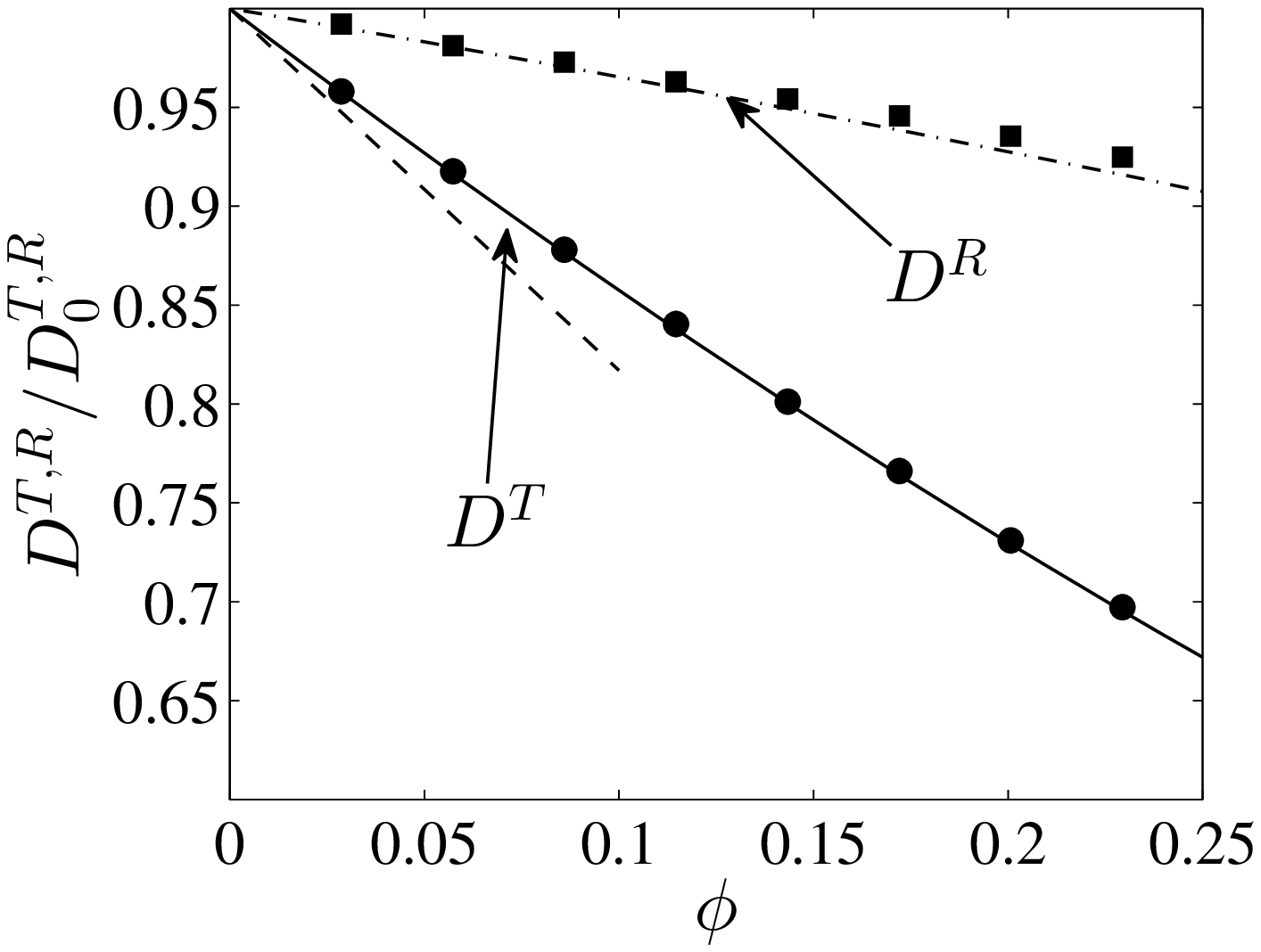}}
	\caption{(a) Periodic array mobility coefficient, $\gamma$, as a function of $\phi$.  The markers indicate the values given by fluctuating FCM, while the dash lines shows those determined by constant force FCM simulations.  (b) Short-time diffusion coefficients, $D^{T}$ and $D^{R}$.  The markers indicate the fluctuating FCM values, while the the solid and dash--dotted lines show those provided by Banchio \& Brady \cite{Banchio2003}.  The dashed line shows the values of $D^T$ for small $\phi$ found by Batchelor \cite{Batchelor1976}.}
	\label{fig:short_time_diffusion}
	\end{center}
\end{figure}

In the first set of computations, we consider the short-time self-diffusion coefficient,
\begin{equation}
D^T=\frac{k_B T}{3N}\textrm{trace}(\mathcal{M}^{\mathcal{VF}}).
\end{equation}
for a periodic array of spheres and a random suspension.  Based on the relationship between the velocity correlations and the mobility matrix, Eq. (\ref{eq:PM2}), we calculate $D^T$ from fluctuating FCM simulations using
\begin{equation}
D^T = \frac{\Delta t}{6N}\sum_{n=1}^N\left\langle \tilde\mathbf{V}_n\cdot \tilde\mathbf{V}_n\right\rangle.
\label{eq:std_fluc}
\end{equation}

\subsubsection{Periodic array of spheres}
For a periodic array, the mobility matrix can be written as 
\begin{equation}
\mathcal{M}^{\mathcal{VF}} = \frac{\gamma(\phi)}{6\pi a \eta}\mathcal{I} 
\end{equation}
where the coefficient $\gamma$ depends on the volume fraction, $\phi$, occupied by the array.  The value of $\gamma$ can be determined by considering a single particle settling under a unit force in a triply periodic domain.  This calculation has been performed for FCM \cite{Maxey2001} and compared well up to volume fractions of $\phi = 0.2$ with the theoretical results of Hasimoto \cite{Hasimoto1959} and Sangani \& Acrivos \cite{Sangani1982}.  

By calculating the short-time self-diffusion coefficient using Eq. (\ref{eq:std_fluc}), the coefficient $\gamma$ can also be determined from the random particle velocities given by fluctuating FCM.  In these simulations, we set $N=1$ and compute the velocity of a sphere located at the centre of the domain.  The force, torque, and stresslet on the particle are set to zero.  For each volume fraction, we determine the particle's velocity for $10^4$ realisations of the fluctuating stress field and average over these realisations to determine $D^T$ and $\gamma$.  

Fig. \ref{fig:lattice} shows the values of $\gamma$ given by fluctuating FCM along with those found by allowing the particle to settle under a constant force.  The values of $\gamma$ given by both approaches are nearly identical over the entire range of $\phi$, confirming our theoretical analysis presented in the previous sections.  Their agreement also indicates that our numerical implementation of fluctuating FCM does indeed give particle velocity statistics that correspond to the FCM mobility matrix.

\subsection{Short-time self-diffusion of a random suspension}

To calculate the short-time self-diffusion coefficient for a suspension, we perform fluctuating FCM simulations with the particles randomly distributed in the domain.  These calculations are performed for $N=50 - 400$, corresponding to the range of volume fractions $\phi = 0.0285 - 0.23$.  We set the forces and torques on the particles to be zero, however, we include the particle stresslets in the computations.  For each volume fraction, we compute the particle velocities for $10^4$ realisations of the fluctuating stress field and average over them to find the short-time self-diffusion coefficient according to Eq. (\ref{eq:std_fluc}).

To compare with previous results, we must correct for the periodicity of the domain using the following relation \cite{Ladd1990,Banchio2003}
\begin{equation}
D^T = D^T_{PER}+\frac{k_BT}{6\pi a\bar{\eta}}(1.7601(\phi/N)^{1/3}-\phi/N)
\end{equation}
where $D^T_{PER}$ is the short-time self-diffusion coefficient for the periodic domain and $\bar{\eta}$ is the suspension viscosity that we determine from independent FCM simulations.  The corrected values of $D^T$ given by fluctuating FCM are shown in Fig. \ref{fig:std}.  The values are normalised by $D^T_0 = k_BT/(6\pi a \eta)$.  For comparison, the solid line in Fig. \ref{fig:std} show results from far-field Stokesian Dynamics calculations \cite{Banchio2003} where it was found that $D^T/D^T_0 \approx 1 - 1.5\phi + 0.75\phi^2$.  We see that fluctuating FCM reproduces this dependence quite well, indicating the changes in mobility due to the stresslets are captured in our simulations.  The dashed line in Fig.  \ref{fig:std} shows the low volume fraction, short-time self-diffusion coefficient, $D^T/D^T_0 = 1 - 1.83\phi + O(\phi^2)$  calculated by Batchelor \cite{Batchelor1976}.  These values are based on exact, two-body hydrodynamics and include the near-field lubrication effects that are neglected in fluctuating FCM and far-field Stokesian Dynamics.  Thus, to recover this asymptotic result, the near-field corrections would also need to be included in fluctuating FCM.  

In addition to $D^T$, we determine the short-time rotational self-diffusion coefficient by calculating
\begin{equation}
D^R = \frac{\Delta t}{6N}\sum_{n=1}^N\left\langle \tilde{\bm{\Omega}}_n\cdot \tilde{\bm{\Omega}}_n\right\rangle.
\end{equation}
The fluctuating FCM values of $D^R$ normalised by $D^R_0 = k_BT/(8 \pi a^3 \eta)$ are also shown in Fig. \ref{fig:short_time_diffusion}.  We again compare our values with the far-field Stokesian Dynamics results taken from \cite{Banchio2003} where it was found that $D^R/D^R_0 \approx 1 - 0.33\phi - 0.16\phi^2$.  We see that the volume fraction dependence of $D^R$ given by fluctuating FCM closely matches that given by far-field Stokesian dynamics.  There is a slight difference in these data, which, after performing additional simulations using different domain sizes, we may attribute to the effects of periodicity.  

\subsection{Long-time self-diffusion of interacting particles}
\begin{figure}
	\begin{center}
	\subfigure[\label{fig:yuk_traj}]{\includegraphics[width=16.25pc]{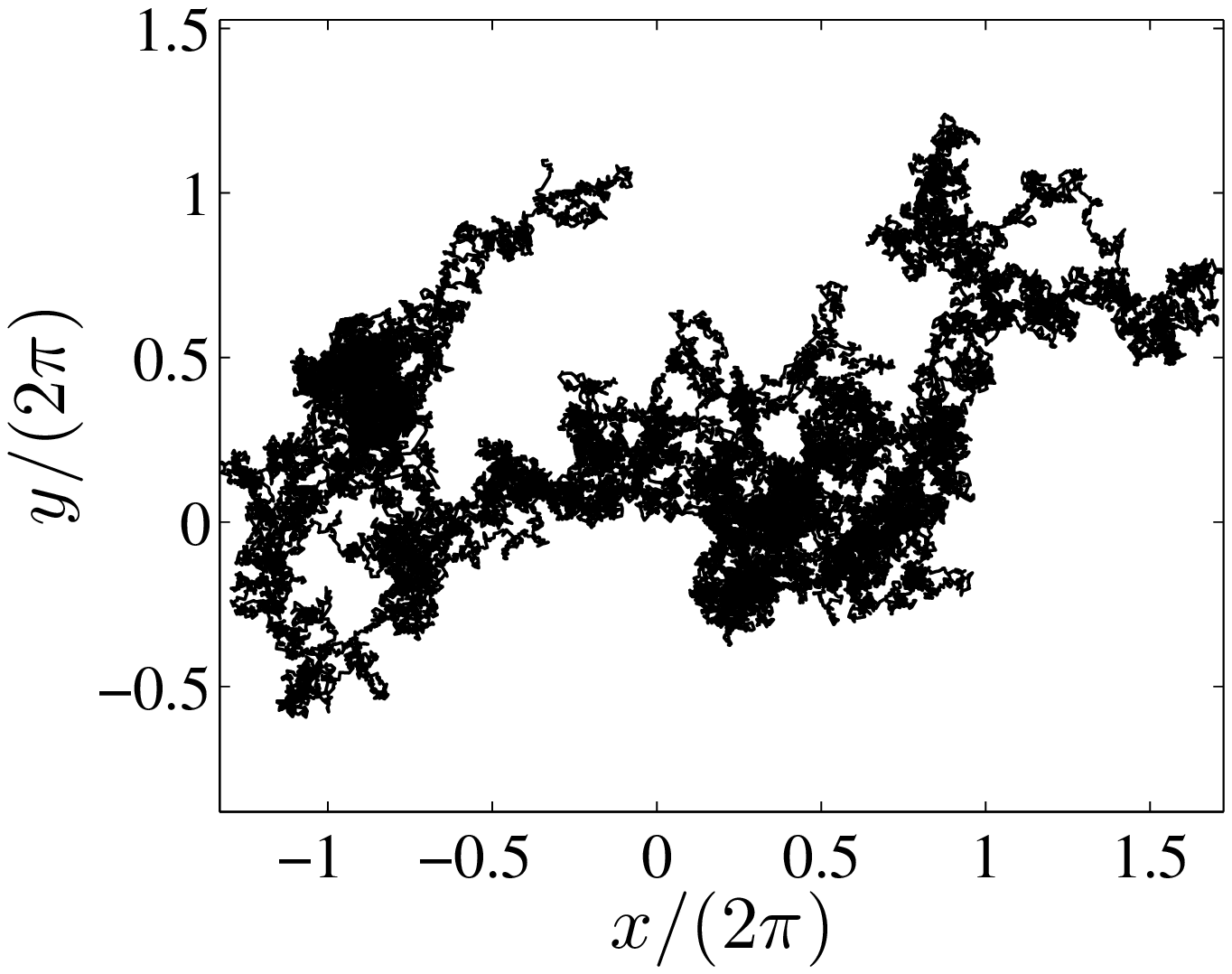}}
	\subfigure[\label{fig:ltmsd}]{\includegraphics[width=16.25pc]{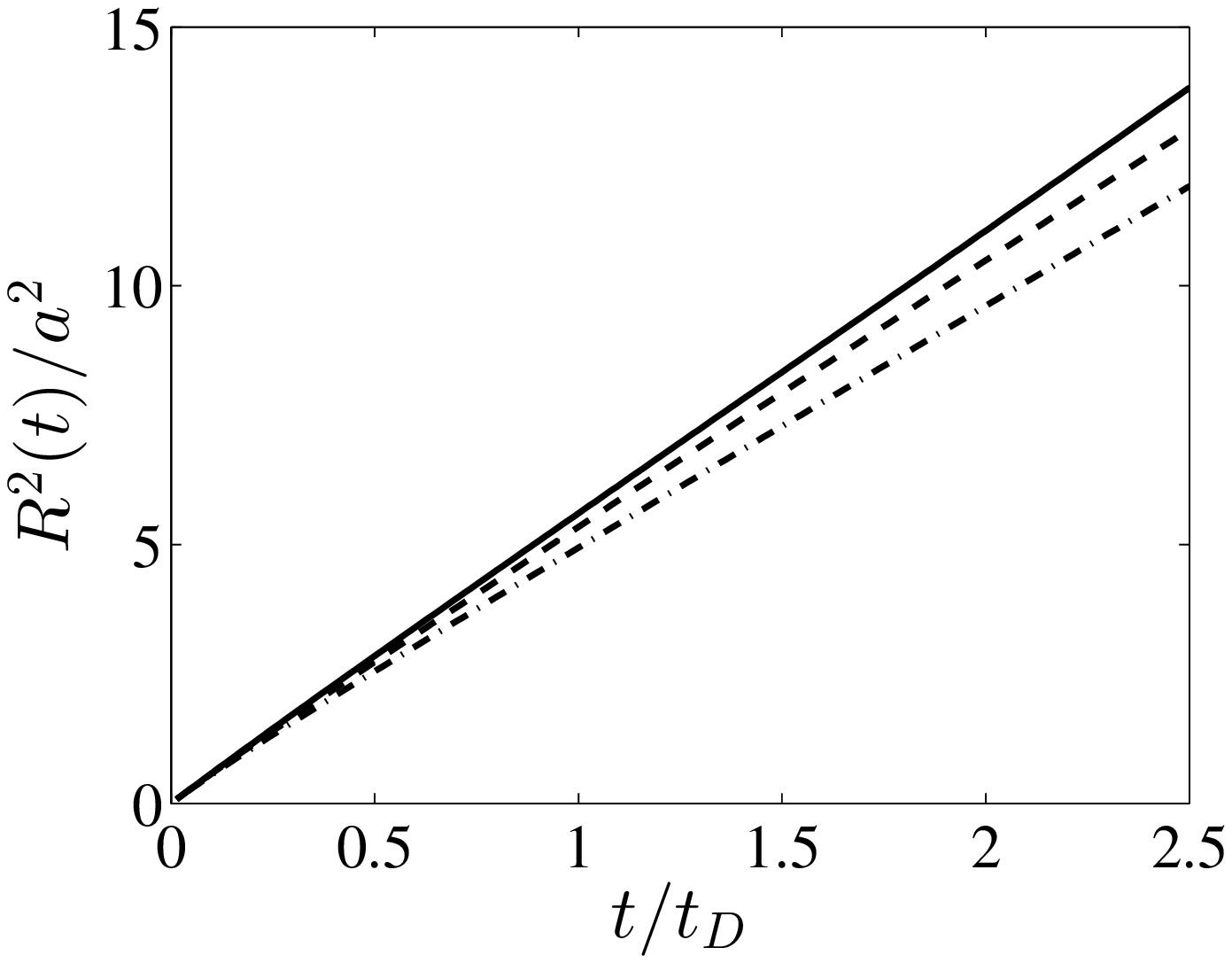}}
	\caption{(a) A particle's trajectory in the $xy$-plane from a stresslet-free fluctuating FCM simulation for which the particles interact via the Yukawa potential and $n\sigma^3_Y = 0.3$ $(\phi=0.15)$.  (b) Mean-squared displacement, Eq. (\ref{eq:MSD}), versus time from stresslet-free fluctuating FCM simulations with Yukawa interactions.  The solid line corresponds to $n\sigma^3_Y = 0.2$, the dashed line $n\sigma^3_Y = 0.3$, while the dash--dotted line shows the mean-squared displacement for $n\sigma^3_Y = 0.4$.}
	\label{fig:long_time_msd}
	\end{center}
\end{figure}

\begin{table}
\caption{Long-time diffusion coefficients for Brownian suspensions with Yukawa interactions as given by the Brownian dynamics simulations of L\"{o}wen \& Szamel \cite{Lowen1993} and fluctuating FCM for the different values of $n\sigma^3_Y$.}
\label{tab:ltd}
\vspace*{.1in}
	\begin{center}
		\begin{tabular}{|c|c|c|}
		\hline
		 Simulation & $n\sigma_{Y}^3$ & $D_{\infty}/D_0$ \\
		\hline \hline
		L\"{o}wen \& Szamel \cite{Lowen1993}    & 0.2  & 0.88(4) \\
		\hline
		 & 0.3 & 0.77(2) \\
		\hline
		& 0.4 & 0.68(2) \\
		\hline \hline
		FCM  & 0.2  & 0.934 $\pm$ 4e-3  \\
		\hline
		& 0.3  & 0.8379 $\pm$ 8e-4  \\
		\hline
		& 0.4  & 0.7698 $\pm$ 8e-4  \\
		\hline \hline
		FCM with stresslets  & 0.2  & 0.805 $\pm$ 3e-3  \\
		\hline
		& 0.3  & 0.7151 $\pm$ 4e-4  \\
		\hline
		& 0.4  & 0.6363 $\pm$ 4e-4  \\
		\hline
		\end{tabular}
	\end{center}
\end{table}
In addition to hydrodynamic interactions, colloidal particles in suspension can interact via a range of other mechanisms such as electrostatic and surface forces.  These additional interactions can further modify diffusive behaviour.  Here, we study these effects using fluctuating FCM to calculate the long-time self-diffusion coefficient
\begin{equation}
D_\infty = \lim_{t \rightarrow \infty} \frac{R^2(t)}{6t}
\label{eq:Dinf}
\end{equation}
from the mean-squared displacement
\begin{equation}
R^2(t) = \frac {1}{N}\sum_{n} \langle |\mathbf{Y}_{n}(t) - \mathbf{Y}_n(0)|^2 \rangle
\label{eq:MSD}
\end{equation}
for a suspension of particles interacting via the soft, pairwise screened Coulomb, or Yukawa potential 
\begin{equation}
V(r) = \frac{U_0 \sigma_{Y}}{r} \exp\left(-\lambda(r - \sigma_{Y})/\sigma_{Y}\right)
\label{eq:yuk}
\end{equation}
given by DLVO theory \cite{Verwey1948,Russel1989}.  The Yukawa potential models the electrostatic repulsion between similarly charged colloidal particles when ions are present in the surrounding fluid.  In Eq. (\ref{eq:yuk}), the strength of the repulsion is set by $U_0$, $\sigma_{Y}$ represents the diameter of the particle, and $\sigma_{Y}/\lambda$ provides the Debye length, the distance over which the electrostatic interactions are screened by the ions.

Similar simulations have been performed using Brownian dynamics \cite{Lowen1993}, however, in these simulations, the hydrodynamic interactions between the particles were ignored.  By comparing with these previous results, we can illustrate the effects of hydrodynamic interactions on particle diffusion in these dispersions.  We, therefore, in Eq. (\ref{eq:yuk}) take the same parameter values as \cite{Lowen1993}, where $U_0 = k_B T$ and the dimensionless screening parameter $\lambda = 8$.  We perform these simulations for volume fractions $\phi = 0.10, 0.15$, and $0.2$, corresponding to $N = 183$, $N = 275$, and $N = 366$, respectively.  Taking $\sigma_Y = 2a$, these values of $\phi$ correspond to $n \sigma_Y^3= 0.2$, $0.3$, and $0.4$ in \cite{Lowen1993}.  For each case, we perform fluctuating FCM simulations with and without the stresslets.  The simulations run for a total time $t = 110t_D$ with time step $\Delta t =$ 0.0013$t_D$.  The timescale $t_D = a^2/D_0$ is based on the short-time diffusion coefficient for a single particle in the periodic domain, $D_0 = 0.854 k_BT/(6\pi a \eta)$.    An example particle trajectory from the $n \sigma_Y^3= 0.3$ simulation is shown in Fig. \ref{fig:yuk_traj}.  From these trajectories, we compute the mean-squared displacement using the particle positions for $t  \geq 10t_D$.  The values of $R^2$ as a function of time for the stresslet-free simulations are shown in Fig. \ref{fig:ltmsd}.  For each case, we observe the linear dependence of $R^2$ on $t$ that is characteristic of diffusive behaviour.  We determine $D_\infty$ by finding the slopes of these lines, which we see decrease as $n\sigma^3_Y$ (and $\phi$) increases.  The values of $D_\infty$ from the fluctuating FCM simulations, as well as those from \cite{Lowen1993} are shown in Table \ref{tab:ltd}.  We see that for each case, $D_\infty$ decreases as $n\sigma^3_Y$ increases.  We also see that the values of $D_\infty$ given by fluctuating FCM without the stresslets, but where hydrodynamic interactions are still present, are greater than those from \cite{Lowen1993}.  Similar enhancements in long-time diffusion due to hydrodynamic interactions have been found previously in simulations \cite{Zahn1997}, and later were confirmed by comparison with experiments \cite{Rinn1999}.  We see, however, that when the stresslets are included, that that the values of $D_{\infty}/D_0$ do decrease dramatically.  This decrease is presumably linked to the lower values of the short-time self-diffusion coefficients (see Fig. \ref{fig:std}) observed when the stresslets are included.

\subsection{Concentration profiles in an external potential}
\begin{figure}
	\begin{center}
	\subfigure[\label{fig:c_noyuk}]{\includegraphics[width=16.25pc]{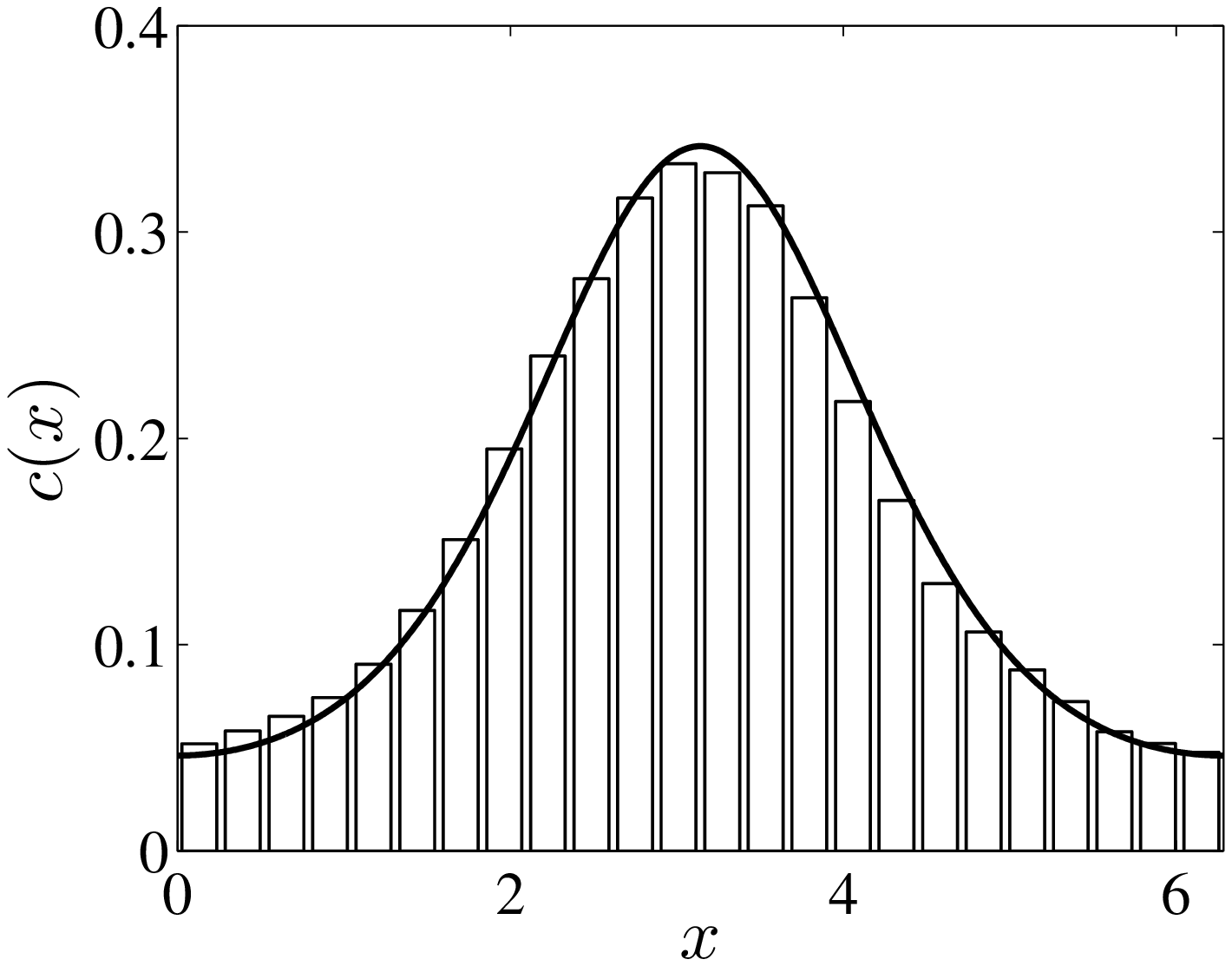}}
	\subfigure[\label{fig:c_yuk}]{\includegraphics[width=16.25pc]{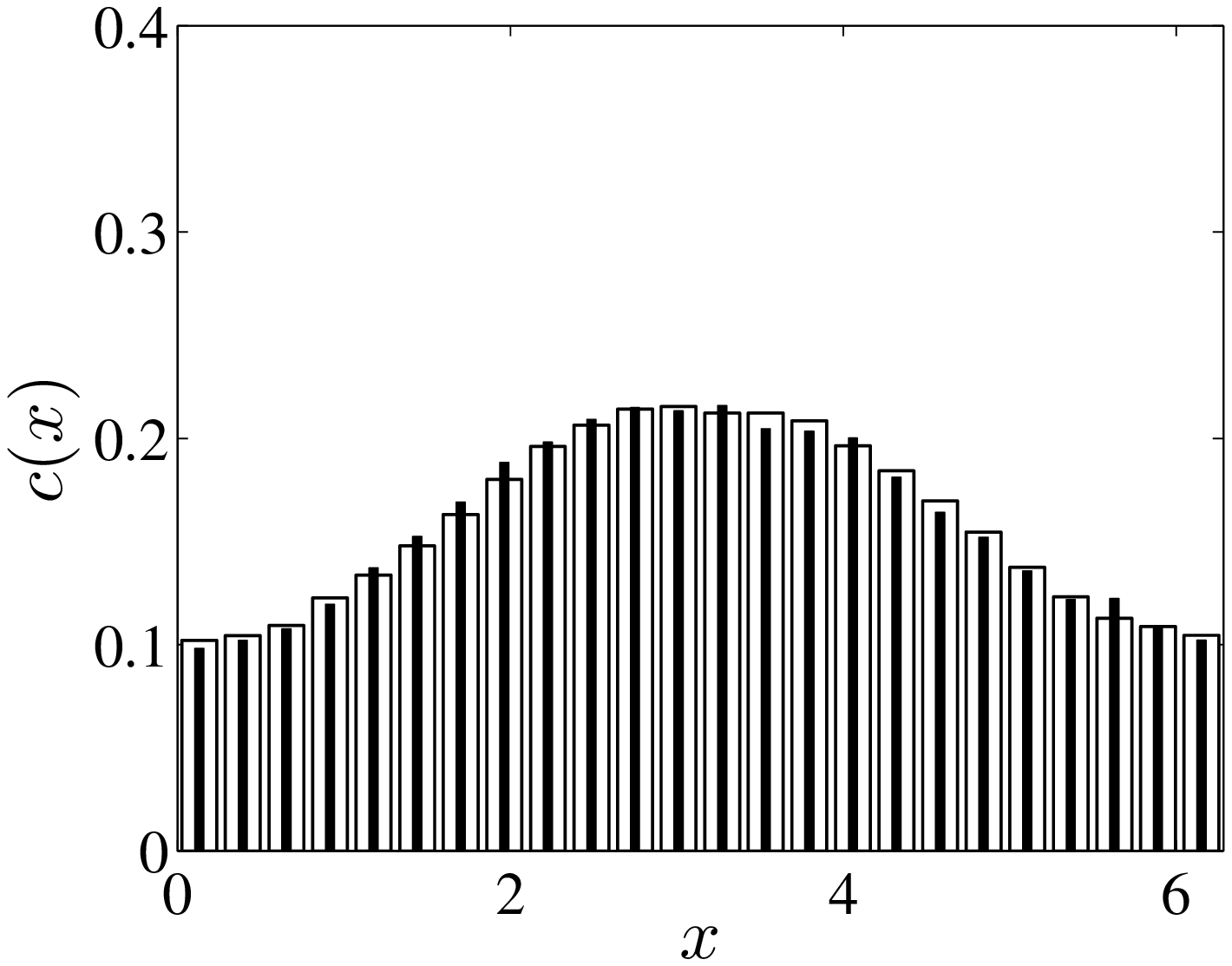}}
	\caption{Concentration profiles for a suspension subject to the external potential $\Phi(x) = \Phi_0\cos x$ (a) The solid line shows the Boltzmann distribution, see Eq. (\ref{eq:Boltzmann}), while the bars show the concentration given by stresslet-free fluctuating FCM with no particle interactions.  (b) Concentration profiles given by fluctuating FCM simulations with Yukawa interactions.  The open bars correspond to stresslet-free simulations, while the closed bars show results from simulations where the particle stresslets are included.}
	\label{fig:short_time_diffusion}
	\end{center}
\end{figure}

In this set of simulations, we consider a suspension of particles subject to the periodic external potential
\begin{equation}
\Phi(x) = \Phi_0 \cos x.
\end{equation}
For non-interacting particles, the equilibrium concentration profile will be given by the Boltzmann distribution 
\begin{equation}
c(x) = \frac{1}{Z} \exp \left( -\Phi_0 \cos x/k_BT \right)
\label{eq:Boltzmann}
\end{equation}
where $Z = \int_0^{2\pi} \exp \left(-\Phi_0 \cos x/k_BT\right) dx.$  We performed stresslet-free fluctuating FCM simulations with $N = 183$ and with the force on particle $n$ given by
\begin{equation}
\mathbf{F}_n = -\frac{d\Phi}{dx} \Big |_{x = X_n} \hat\mathbf{x}
\end{equation}
where $X_n = \mathbf{Y}_n \cdot \hat{\mathbf{x}}$.  The simulation is run to $t = 130t_D$ with a time step of $t = 0.0013 t_D$.  Fig. \ref{fig:c_noyuk} shows the time-averaged concentration for this simulation.  We see that fluctuating FCM reproduces quite well the equilibrium concentration given by Eq. (\ref{eq:Boltzmann}).  This, however, changes when we allow for interactions between the particles.  We perform the same simulation, but now include particle interactions via the Yukawa potential with the same parameters used in the long-time diffusion simulations.  The resulting concentration profiles for simulations with and without the stresslets are shown in Fig. \ref{fig:c_yuk}.  We see that hydrodynamic interactions do not affect the equilibrium profile as both simulations yield nearly identical results.  We do see, however, that since the Yukawa interactions modify the total energy of the system, the spatial distribution of particles is modified, and when compared with the case where there are no interactions, Fig. \ref{fig:c_noyuk}, it is closer to being uniform.

\subsection{Enhanced diffusion in cellular flows}
\begin{figure}
	\begin{center}
	\includegraphics[height=3.0in]{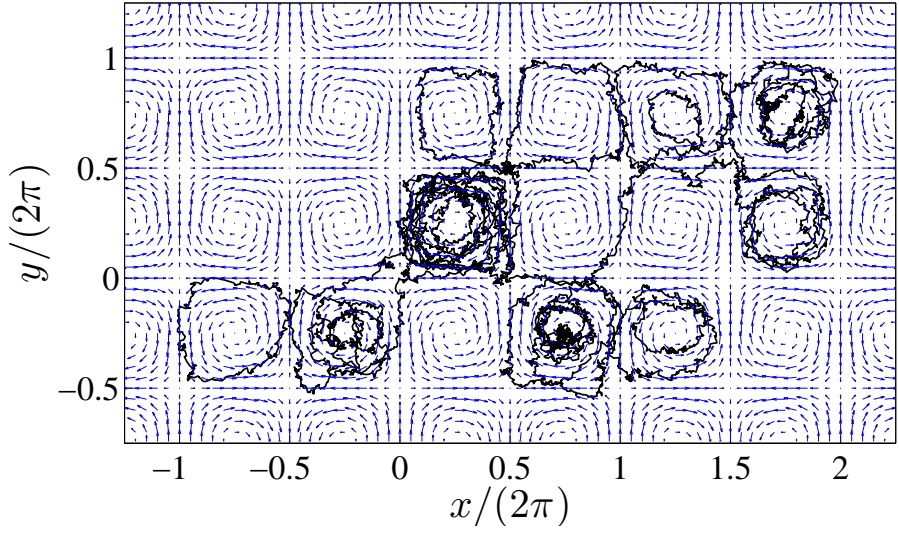}
	\caption{The cellular flow field, Eq. (\ref{eq:cellflow}), along with a particle's trajectory from the fluctuating FCM simulation where the particle stresslets are included.  In the simulation, the particles interact via the Yukawa potential and $n\sigma_Y^3 = 0.2$ $(\phi=0.10)$.}
	\label{fig:flow_traj}
	\end{center}
\end{figure}

\begin{figure}
	\begin{center}
	\includegraphics[height=3.0in]{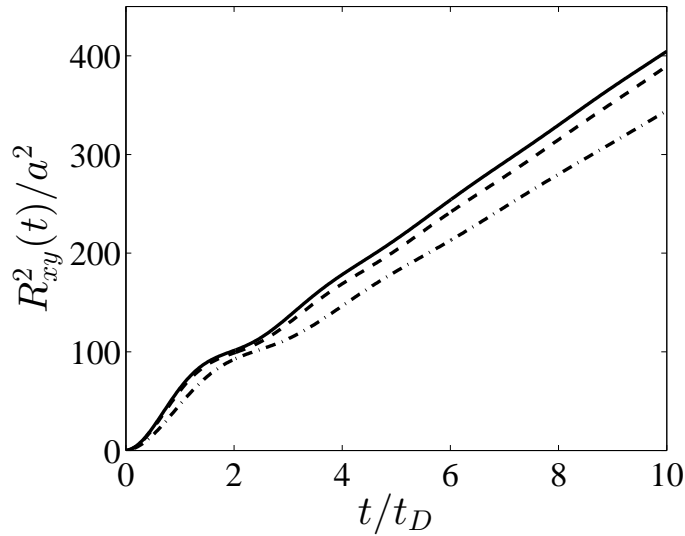}
	\caption{Mean-squared displacement in the $xy$-plane as a function of time.  The solid line corresponds to the stresslet-free fluctuating FCM simulation of non-interacting particles, the dashed line show the values from the stresslet-free fluctuating FCM simulation with Yukawa interactions, and the dash-dotted shows the results from the fluctuating FCM simulation that includes both particle stresslets and Yukawa interactions.}
	\label{fig:flow_msd}
	\end{center}
\end{figure}

As a final numerical example, we consider a suspension of particles in a periodic, cellular flow field (see Fig. \ref{fig:flow_traj})
\begin{equation}
\mathbf{u}_{cell}(\mathbf{x}) = \frac{\alpha}{2\eta}\left(-\sin x \cos y \hat{\mathbf{x}} + \cos x \sin y \hat{\mathbf{y}}\right).
\label{eq:cellflow}
\end{equation}
Transport in cellular flow fields has served as a fundamental mathematical model to understand particle motion in turbulent flows, especially for inertial particles \cite{Maxey1986,Maxey1987}, self-propelled particles \cite{Durham2011}, and elastic filaments \cite{Young2007}.  For Brownian tracers, it has been shown through asymptotic analysis and homogenisation of the advection-diffusion equation \cite{Majda1999}, that at long times particle motion becomes diffusive with a diffusion coefficient that is enhanced by the flow.  

We can examine the transport of Brownian particles in cellular flows using fluctuating FCM, where we may also consider the effects of particle interactions.  In the simulations, the cellular flow field is incorporated by including the additional body force 
\begin{equation}
\mathbf{f}_{cell}(x, y) = \alpha \left(-\sin x \cos y \hat{\mathbf{x}} + \cos x \sin y \hat{\mathbf{y}}\right)
\end{equation}
in the Stokes equations.  This new total flow field is used to determine the velocities and angular velocities of the particles, as well as the local rates-of-strain if the stresslets are to be included.  We perform the simulations with $N=183$ and $\alpha=10$ for three separate cases, one without the stresslets or particle interactions, a second with Yukawa interactions, but without the stresslets, and a third with both stresslets and Yukawa interactions.  We run our simulations until $t = 130 t_D$, again with time step $\Delta t = 0.0013 t_D$.  A particle's trajectory from the fluctuating FCM simulation with stresslets and Yukawa interactions is shown in Fig. \ref{fig:flow_traj}.  We see that the particle is carried along by the flow, but Brownian motion allows it to move across streamlines, and eventually go from one cell to another.  Fig. \ref{fig:flow_msd} shows the mean-squared  $xy$-displacement
\begin{equation}
R^2_{xy}(t) = \frac {1}{N}\sum_{n} \langle (X_{n}(t) - X_n(0))^2 + (Y_{n}(t) - Y_n(0))^2\rangle
\label{eq:MSDxy}
\end{equation}
as a function of time for all three simulations.  In each case, we see that at long times, $R^2_{xy}(t)$ depends linearly on $t$, and the motion is diffusive.  From the slope, we can determine the effective long-time diffusion coefficient, $D^{xy}_{\infty}$, for each simulation.  Without stresslets or Yukawa interactions, we find that $D^{xy}_{\infty}/D_0 = 6.29$, when there are only Yukawa interactions we have $D^{xy}_{\infty}/D_0 = 6.07$, and when there are both the Yukawa interactions and stresslets, $D^{xy}_{\infty}/D_0 = 5.42$.  In each case, we see that the diffusion coefficient is much greater than the corresponding values without the cellular flow (see the $n \sigma^3_Y = 0.2$ cases in Table \ref{tab:ltd}) and we find the greatest enhancement ($D^{xy}_{\infty}/D_{\infty} = 6.73$) when both the Yukawa interactions and stresslets are present.  This enhancement is in agreement with the results from \cite{Majda1999}, however, we see also that in suspensions of interacting particles, the enhancement is, in fact, magnified.

\section{Conclusions}
In this paper, we presented fluctuating FCM and demonstrated its effectiveness as an approach to simulate the dynamics of dilute colloidal suspensions.  This method involves computing the fluid flows generated by a fluctuating stress and employing the FCM framework to determine particle Brownian motion.  We have shown analytically that fluctuating FCM yields random particle velocities and angular velocities with correlations consistent with the fluctuation-dissipation theorem even when higher-order multipoles, i.e. the stresslets, are used.  In addition, we showed that for dynamic simulations, Brownian drift can be resolved using the midpoint time integration scheme developed by Fixman \cite{Fixman1978,Grassia1995} and the conjugate gradient method to obtain the Brownian forces and torques.  We have conducted several numerical experiments confirming our theoretical results, demonstrating that fluctuating FCM yields the correct diffusion for hydrodynamically interacting particles.  We have also shown the method's versatility and how particle interactions can affect diffusion coefficients, suspension concentration profiles in external potentials, and enhanced diffusion in external flow fields.  

There are several directions in which fluctuating FCM can be extended, or modified to be used with other schemes.  In our theoretical analysis, we did not use the specific properties (other than differentiability) of the Gaussian envelopes that regularise the multipole expansion and volume average the flow field.  Thus, the flows generated by fluctuating stresses could also be readily integrated with other regularisation schemes, such as the method of regularised Stokeslets \cite{Cortez2001,Cortez2005}.  Additionally, it might be possible to use random flows in conjunction with particle-mesh Ewald schemes \cite{Saintillan2005,HernandezOrtiz2007}.  As we saw in examining the short-time diffusion coefficient, lubrication and near-field hydrodynamic interactions can affect the properties of Brownian suspensions, even at low volume fractions.  We are currently investigating how to incorporate lubrication effects \cite{Dance2003,Yeo2010} into fluctuating FCM and enable the large-scale simulation of dense Brownian suspensions.  Another important direction is the incorporation of particle and/or fluid inertia into fluctuating FCM.  It has been demonstrated \cite{Hinch1975,Atzberger2007,Balboa2013} that inertia can lead to power-law time correlations and it would also be of interest to explore these effects in large-scale suspension simulations.  

\section*{Acknowledgments}
I wish to thank Martin Maxey, Michael Shelley, and Kyongmin Yeo for valuable discussions during the course of this work and Aleksandar Donev for helpful comments on the original manuscript.  I also acknowledge support from the EPSRC Small Equipment Funding Scheme for Early Career Researchers under grant EP/K030760/1.

\appendix
\section{Force-coupling method mobility matrices}\label{sec:FCMapp}
In Section \ref{sec:FCM}, we showed that
\begin{eqnarray}
\mathcal{M}^{\mathcal{V}\mathcal{F}}_{FCM; n m} &=&\int \int \Delta_n(\mathbf{x}) \mathbf{G}(\mathbf{x} - \mathbf{y}) \Delta_m(\mathbf{y})d^3\mathbf{x}d^3\mathbf{y}.
\label{eq:mvf}
\end{eqnarray}
One can obtain similar expressions for the other submatrices of the FCM grand mobility matrix (Eq. (\ref{eq:FCM5})).  This is done by first writing in terms of the Stokeslet, $\mathbf{G}$, the flows generated by the FCM force distributions corresponding to the force, torque, and stresslet of particle $m$, then, using Eqs. (\ref{eq:FCM3a} -- \ref{eq:FCM3c}), showing how they contribute to the velocity, angular velocity, and local rate-of-strain of particle $n$.  Where appropriate, integration by parts can be used to move the partial derivatives on $\mathbf{G}$ onto the FCM Gaussian envelopes.  The mobility matrices, especially those related to the rate-of-strain, are most conveniently written using index notation.  We use this notation here, and in doing so, we remove the subscript label ``FCM'' for clarity. 

We first consider the flow generated when there is a force on particle $m$, see Eq. (\ref{eq:fcmfu}).  If we take this force to be of unit magnitude and in the $j$ direction, we find that the submatrix entry corresponding to the angular velocity of particle $n$ in the $i$ direction is  
\begin{equation}
\mathcal{M}^{\mathcal{W}\mathcal{F}}_{n,i; m,j} = \frac{1}{2}\int \int \epsilon_{ikl}\frac{\partial \Theta_n(\mathbf{x})}{\partial x_l}G_{kj}(\mathbf{x} - \mathbf{y}) \Delta_n(\mathbf{y})d^3\mathbf{x}d^3\mathbf{y}, 
\label{eq:fcmmwf}
\end{equation}
If we consider instead the $ik$ entry of the local rate-of-strain for particle $n$, we find that
\begin{eqnarray}
\mathcal{M}^{\mathcal{E}\mathcal{F}}_{n,ik; m,j} &=& -\frac{1}{2}\int \int\frac{\partial \Theta_n(\mathbf{x})}{\partial x_k}G_{ij}(\mathbf{x} - \mathbf{y}) \Delta_n(\mathbf{y})d^3\mathbf{x}d^3\mathbf{y} \nonumber \\
&& - \frac{1}{2}\int \int \frac{\partial \Theta_n(\mathbf{x})}{\partial x_i}G_{kj}(\mathbf{x} - \mathbf{y}) \Delta_n(\mathbf{y})d^3\mathbf{x}d^3\mathbf{y}.
\end{eqnarray}

When the fluid velocity is a result of a unit torque on particle $m$ in the $j$ direction, the entry of the submatrix is  
\begin{equation}
\mathcal{M}^{\mathcal{W}\mathcal{T}}_{n,i;m,j} =\frac{1}{4}\int \int \epsilon_{ipq}\frac{\partial \Theta_n(\mathbf{x})}{\partial x_q} G_{pk}(\mathbf{x} - \mathbf{y}) \epsilon_{klj}\frac{\partial \Theta_m(\mathbf{y})}{\partial y_l}d^3\mathbf{x}d^3\mathbf{y}
\end{equation}
for the angular velocity of particle $n$ in the $i$ direction, while we have
\begin{eqnarray}
\mathcal{M}^{\mathcal{E}\mathcal{T}}_{n,ik; m,j} &=& -\frac{1}{2}\int \int\frac{\partial \Theta_n(\mathbf{x})}{\partial x_k}G_{ip}(\mathbf{x} - \mathbf{y}) \epsilon_{plj}\frac{\partial \Theta_m(\mathbf{y})}{\partial y_l}d^3\mathbf{x}d^3\mathbf{y} \nonumber \\
&&- \frac{1}{2}\int \int \frac{\partial \Theta_n(\mathbf{x})}{\partial x_i}G_{kp}(\mathbf{x} - \mathbf{y})\epsilon_{plj}\frac{\partial \Theta_m(\mathbf{y})}{\partial y_l}d^3\mathbf{x}d^3\mathbf{y}
\end{eqnarray}
for the $ik$ entry of the local rate-of-strain of particle $n$.  It can also be shown that $\mathcal{M}^{\mathcal{W}\mathcal{F}}_{n,i; m,j} = \mathcal{M}^{\mathcal{V}\mathcal{T}}_{m,j; n,i}$, which gives $\mathcal{M}^{\mathcal{W}\mathcal{F}}_{FCM} = (\mathcal{M}^{\mathcal{V}\mathcal{T}}_{FCM})^{T}$.

Finally, considering the flow generated by the $kl$ entry of the stresslet on particle $m$ and the $ij$ component of the local rate-of-strain on particle $n$, we have that
\begin{eqnarray}
\mathcal{M}^{\mathcal{E}\mathcal{S}}_{n,ij; m,kl} &=&\frac{1}{4}\int \int \frac{\partial \Theta_n(\mathbf{x})}{\partial x_j} G_{ik}(\mathbf{x} - \mathbf{y})\frac{\partial \Theta_m(\mathbf{y})}{\partial y_l}d^3\mathbf{x}d^3\mathbf{y} \nonumber\\
&&+\frac{1}{4}\int \int \frac{\partial \Theta_n(\mathbf{x})}{\partial x_i} G_{jk}(\mathbf{x} - \mathbf{y})\frac{\partial \Theta_m(\mathbf{y})}{\partial y_l}d^3\mathbf{x}d^3\mathbf{y} \nonumber\\
&&+\frac{1}{4}\int \int \frac{\partial \Theta_n(\mathbf{x})}{\partial x_j} G_{il}(\mathbf{x} - \mathbf{y})\frac{\partial \Theta_m(\mathbf{y})}{\partial y_k}d^3\mathbf{x}d^3\mathbf{y} \nonumber\\
&&+\frac{1}{4}\int \int \frac{\partial \Theta_n(\mathbf{x})}{\partial x_i} G_{jl}(\mathbf{x} - \mathbf{y})\frac{\partial \Theta_m(\mathbf{y})}{\partial y_k}d^3\mathbf{x}d^3\mathbf{y}. 
\end{eqnarray}
By the symmetry of the grand mobility matrix, the remaining submatrices are related to those already determined \cite{Yeo2010}.  Specifically, we have $\mathcal{M}^{\mathcal{E}\mathcal{F}}_{FCM} = -(\mathcal{M}^{\mathcal{V}\mathcal{S}}_{FCM})^T$ and $\mathcal{M}^{\mathcal{E}\mathcal{T}}_{FCM}= -(\mathcal{M}^{\mathcal{W}\mathcal{S}}_{FCM})^T$.

\section{Flow statistics due to a fluctuating stress}\label{sec:flucstress}

Here, we establish the fluid velocity correlations when the fluid is forced by the fluctuating stress, $\mathbf{P}$.  Recall that $\mathbf{P}$, in index notation, has the following statistics
\begin{eqnarray}
	\left\langle P_{jl}\right\rangle&=&0\\
	\left\langle P_{jl}(\mathbf{x},t)P_{pq}(\mathbf{x}',t') \right\rangle&=&2k_BT\eta\left(\delta_{jp}\delta_{lq} + \delta_{jq}\delta_{lp}\right)\delta(\mathbf{x}-\mathbf{x}')\delta(t-t') 
	\label{eq:Scovar}.
\end{eqnarray}
In Fourier space, the correlation relation, Eq. (\ref{eq:Scovar}), will be 
\begin{eqnarray}
	\left\langle \hat{P}_{jl}(\mathbf{k},t)\hat{P}_{pq}(-\mathbf{k},t') \right\rangle&=&2k_BT\eta\left(\delta_{jp}\delta_{lq} + \delta_{jq}\delta_{lp}\right)\delta(t-t').
	\label{eq:fouriercor}
\end{eqnarray}
We can find the random fluid flow, $\tilde{\mathbf{u}}$, due to $\mathbf{P}$ by solving the Stokes equations, Eq. (\ref{eq:RanVel5}) in Fourier space.  Working in index notation, we find that
\begin{equation}
\hat{\tilde{u}}_j(\mathbf{k},t) = \frac{1}{\eta k^2}\left(\delta_{jm}-\frac{k_jk_m}{k^2}\right)ik_l \hat P_{ml}(\mathbf{k},t),
\end{equation}
and consequently, the correlations of the flow field will be given by
 \begin{eqnarray}
\left\langle \hat{\tilde{u}}_j(\mathbf{k},t)\hat{\tilde{u}}_p(-\mathbf{k},t')\right\rangle &= &\frac{1}{\eta^2 k^4}\left(\delta_{jm}-\frac{k_jk_m}{k^2}\right) \nonumber\\
&& \times \left(\delta_{pn}-\frac{k_pk_n}{k^2}\right)k_lk_q\left\langle \hat P_{ml}(\mathbf{k},t)\hat P_{nq}(-\mathbf{k},t')\right\rangle.
\end{eqnarray}
Substituting Eq. (\ref{eq:fouriercor}) for the stress correlations, we have
 \begin{eqnarray}
\left\langle \hat{\tilde{u}}_j(\mathbf{k},t)\hat{\tilde{u}}_p(-\mathbf{k},t')\right\rangle&=&\frac{2k_BT}{\eta k^4}\left(\delta_{jm}-\frac{k_jk_m}{k^2}\right)\left(\delta_{pn}-\frac{k_pk_n}{k^2}\right)\nonumber\\
& &\times k_lk_q\left(\delta_{mn}\delta_{lq}+\delta_{mq}\delta_{ln}\right)\delta(t-t')
\end{eqnarray}
which further becomes
 \begin{eqnarray}
\left\langle \hat{\tilde{u}}_j(\mathbf{k},t)\hat{\tilde{u}}_p(-\mathbf{k},t')\right\rangle&=&\frac{2k_BT}{\eta k^4}\left(\delta_{jm}\delta_{pn}-\frac{k_jk_m}{k^2}\delta_{pn}-\frac{k_pk_n}{k^2}\delta_{jm}+\frac{k_jk_mk_pk_n}{k^4} \right)\nonumber\\
&&\times \left(\delta_{mn}k^2+k_{m}k_{n}\right)\delta(t-t').
\end{eqnarray}
After expanding and cancelling terms, one determines
 \begin{equation}
\left\langle \hat{\tilde{u}}_j(\mathbf{k},t)\hat{\tilde{u}}_p(-\mathbf{k},t')\right\rangle = \frac{2k_BT}{\eta k^2}\left(\delta_{jp}-\frac{k_jk_p}{k^2}\right)\delta(t-t')
\end{equation}
which in real space is
 \begin{equation}
\left\langle \tilde{u}_j(\mathbf{x},t)\tilde{u}_p(\mathbf{x}',t')\right\rangle = \frac{2k_BT}{8\pi\eta r}\left(\delta_{jp}+\frac{(\mathbf{x}-\mathbf{x}')_j(\mathbf{x}-\mathbf{x}')_p}{r^2}\right)\delta(t-t'),
\label{eq:ucor}
\end{equation}
or
\begin{equation}
\left\langle \tilde{\mathbf{u}}(\mathbf{x},t)\tilde{\mathbf{u}}^{T}(\mathbf{x}',t')\right\rangle = 2k_BT\mathbf{G}(\mathbf{x}-\mathbf{x}')\delta(t-t').
\label{eq:fluidcor_app}
\end{equation}

\section{Fluctuating FCM: Particle velocity correlations}\label{sec:FCMcorr_app}
In this appendix, we show that volume averaging the random flow field, $\tilde\mathbf{u}$, using Eqs. (\ref{eq:FCM3a}) -- (\ref{eq:FCM3c}) gives the random particle velocities, angular velocities, and local rates-of-strain with correlations proportional to the FCM grand mobility matrix.  This calculation was performed in Section \ref{sec:particle_cors_nostress} for the velocity-velocity correlations, and here, we present the correlations for the remaining quantities.  These expressions are fundamental to establishing that fluctuating FCM reproduces the correct statistics for the random motion of the particles.

From Eqs. (\ref{eq:FCM3a}) -- (\ref{eq:FCM3c}), the expressions for the induced velocities, angular velocities, and local rates-of-strain for particle $n$, in index notation, are
\begin{eqnarray}
\tilde{V}_{n,i}&=&\int \tilde{u}_i\Delta_n(\mathbf{x}) d^3\mathbf{x} \nonumber\\
\tilde{\Omega}_{n,i}&=&\frac{1}{2}\int \epsilon_{ijk}\frac{\partial \tilde{u}_k}{\partial x_j}\Theta_n(\mathbf{x}) d^3\mathbf{x} \nonumber\\
\tilde{E}_{n,ij} &=& \frac{1}{2}\int\left(\frac{\partial \tilde{u}_i}{\partial x_j} + \frac{\partial \tilde{u}_j}{\partial x_i}\right)  \Theta^n(\mathbf{x}) d^3\mathbf{x}.
\label{eq:ran_pm}
\end{eqnarray}
Having already established the velocity correlations in Section \ref{sec:particle_cors_nostress}, we now seek the correlations between the velocity of particle $n$ and the angular velocity of particle $m$.  Multiplying these quantities together, integrating by parts, and taking the ensemble average gives us
\begin{eqnarray}
\langle \tilde{V}_{n,i}(t) \tilde{\Omega}_{m,j}(t')\rangle = \frac{1}{2}\int \int \langle\tilde{u}_i(\mathbf{x},t)\tilde{u}_k(\mathbf{y},t')\rangle \epsilon_{klj} \frac{\partial \Theta_{m}}{\partial y_l}\Delta_n(\mathbf{x})d^3\mathbf{x} d^3\mathbf{y}.
\end{eqnarray}
Upon substituting Eq. (\ref{eq:fluidcor_app}) for the fluid flow correlations, we see that
\begin{eqnarray}
\langle \tilde{V}_{n,i}(t) \tilde{\Omega}_{m,j}(t')\rangle = k_BT\delta(t-t')\int \int G_{ik}(\mathbf{x}-\mathbf{y}) \epsilon_{klj} \frac{\partial \Theta_{m}}{\partial y_l}\Delta_n(\mathbf{x})d^3\mathbf{x} d^3\mathbf{y}.\nonumber\\
\end{eqnarray}
From Eq. (\ref{eq:fcmmwf}) and its symmetry properties, we recognise that the right hand side may also be written as
\begin{eqnarray}
\langle \tilde{V}_{n,i}(t) \tilde{\Omega}_{m,j}(t')\rangle = 2k_BT\delta(t-t')\mathcal{M}^{\mathcal{V}\mathcal{T}}_{m,j; n,i},
\end{eqnarray}
which for all particles becomes
\begin{equation}
\langle \tilde{\mathcal{V}}(t) \tilde{\mathcal{W}}^T(t') \rangle = 2k_BT\delta(t-t')\mathcal{M}^{\mathcal{VT}}_{FCM}.
\end{equation}
Repeating the same calculation for the other possible combinations of the quantities in Eq. (\ref{eq:ran_pm}), we can find the remaining correlations
\begin{eqnarray}
\langle \tilde{\Omega}_{n,i}(t) \tilde{\Omega}_{m,j}(t')\rangle&=&\frac{1}{4}\int \int \langle\tilde{u}_p(\mathbf{x},t)\tilde{u}_k(\mathbf{y},t')\rangle \epsilon_{pqi}\epsilon_{klj} \frac{\partial \Theta_{m}}{\partial y_l}\frac{\partial \Theta_{n}}{\partial x_q}d^3\mathbf{x} d^3\mathbf{y} \\
\langle \tilde{V}_{n,i}(t) \tilde{E}_{m,jk}(t')\rangle &=&-\frac{1}{2}\int \int \Delta_n(\mathbf{x})\frac{\partial \Theta_{m}}{\partial y_k} \langle\tilde{u}_i(\mathbf{x},t)\tilde{u}_j(\mathbf{y},t')\rangle d^3\mathbf{x} d^3\mathbf{y} \nonumber \\
&&-\frac{1}{2}\int \int \Delta^n(\mathbf{x})\frac{\partial \Theta_{m}}{\partial y_j} \langle\tilde{u}_i(\mathbf{x},t)\tilde{u}_k(\mathbf{y},t')\rangle. \\ 
\langle \tilde{\Omega}_{n,i}(t) \tilde{E}_{m,jk}(t')\rangle &=&-\frac{1}{4}\int \int  \epsilon_{pqi}\frac{\partial \Theta_{n}}{\partial x_q} \frac{\partial \Theta_{m}}{\partial y_k} \langle\tilde{u}_p(\mathbf{x},t)\tilde{u}_j(\mathbf{y},t')\rangle d^3\mathbf{x} d^3\mathbf{y} \nonumber \\
&&-\frac{1}{4}\int \int \epsilon_{pqi}\frac{\partial \Theta_{n}}{\partial x_q} \frac{\partial \Theta_{m}}{\partial y_j} \langle\tilde{u}_p(\mathbf{x},t)\tilde{u}_k(\mathbf{y},t')\rangle d^3\mathbf{x} d^3\mathbf{y} \\
\langle \tilde{E}_{n,ij}(t) \tilde{E}_{m,kl}(t')\rangle &=&\frac{1}{4}\int \int  \frac{\partial \Theta_{n}}{\partial x_j} \frac{\partial \Theta_{m}}{\partial y_l} \langle\tilde{u}_i(\mathbf{x},t)\tilde{u}_k(\mathbf{y},t') \rangle d^3\mathbf{x} d^3\mathbf{y} \nonumber \\
&&+\frac{1}{4}\int \int \frac{\partial \Theta_{n}}{\partial x_j} \frac{\partial \Theta_{m}}{\partial y_k} \langle\tilde{u}_i(\mathbf{x},t)\tilde{u}_l(\mathbf{y},t')\rangle d^3\mathbf{x} d^3\mathbf{y} \nonumber \\ 
&&+\frac{1}{4}\int \int \frac{\partial \Theta_{n}}{\partial x_i} \frac{\partial \Theta_{m}}{\partial y_l} \langle\tilde{u}_j(\mathbf{x},t)\tilde{u}_k(\mathbf{y},t')\rangle d^3\mathbf{x} d^3\mathbf{y} \nonumber \\
&&+\frac{1}{4}\int \int \frac{\partial \Theta_{n}}{\partial x_i} \frac{\partial \Theta_{m}}{\partial y_k} \langle\tilde{u}_j(\mathbf{x},t)\tilde{u}_l(\mathbf{y},t')\rangle d^3\mathbf{x} d^3\mathbf{y}. 
\end{eqnarray}
After substituting Eq. (\ref{eq:fluidcor_app}) for the fluid velocity correlations, we see that these expressions are entries of the corresponding FCM grand mobility submatrices multiplied by $2k_BT\delta(t-t')$, which, for all particles, become
\begin{eqnarray}
\langle \mathcal{W}(t) \mathcal{W}^T(t') \rangle &=& 2k_BT\delta(t-t')\mathcal{M}^{\mathcal{WT}}_{FCM} \\
\langle \mathcal{V}(t) \mathcal{E}^T(t') \rangle &=& -2k_BT\delta(t-t')\mathcal{M}^{\mathcal{VS}}_{FCM} \\
\langle \mathcal{W}(t) \mathcal{E}^T(t') \rangle &=& -2k_BT\delta(t-t')\mathcal{M}^{\mathcal{WS}}_{FCM} \\
\langle \mathcal{E}(t) \mathcal{E}^T(t') \rangle &=& -2k_BT\delta(t-t')\mathcal{M}^{\mathcal{ES}}_{FCM}
\label{eq:par_cors}
\end{eqnarray}

\bibliographystyle{elsart-num}
\bibliography{BMbib}

\end{document}